\documentclass[
preprint,
tightenlines,
showpacs
,showkeys
,nofootinbib
,aps
,amsfonts
,amssymb
]{revtex4}

\usepackage{graphicx}  
\usepackage{amsmath} 
\usepackage[dvips]{color}
\usepackage[normalem]{ulem}  

\newcommand{\bce}{\begin{center}}  
\newcommand{\ece}{\end{center}}
\newcommand{\beq}{\begin{equation}}  
\newcommand{\eeq}{\end{equation}}
\newcommand{\beqy}{\begin{eqnarray}}
\newcommand{\eeqy}{\end{eqnarray}}

\def\){\right)} 
\def\({\left(} 
\def\]{\right]} 
\def\[{\left[}

\def\Journal#1#2#3#4{{#1} {\bf #2}, #3 (#4)}

\def\PRD{{\em Phys. Rev.} D}

\begin{document}

\title{Viscous damping of $r$-mode oscillations in compact stars with
quark matter}

\author{%
Prashanth Jaikumar\footnote{Electronic address: {\tt jaikumar@imsc.res.in}}}
\affiliation{Institute of Mathematical Sciences, C.I.T Campus, Chennai, 
TN 600113, India}
\affiliation{Physics Division, Argonne National Laboratory, Argonne,
  IL 60439-4843, U.S.A.}

\author{%
Gautam Rupak\footnote{Electronic address: {\tt
    grupak@u.washington.edu}. Permanent address: Department of Physics
and Astronomy, Mississippi State University, Mississippi State, MS 39762}}
\affiliation{Department of Physics, North Carolina State
University, Raleigh, NC 27695, U.S.A.}

\author{%
Andrew W. Steiner\footnote{Electronic address: {\tt steinera@pa.msu.edu}}}

\affiliation{Joint Institute for Nuclear Astrophysics, National
Superconducting Cyclotron Laboratory and \\
Department of Physics
and Astronomy, Michigan State University, East Lansing, MI 48824, U.S.A. }

\begin{abstract} We determine characteristic timescales for the
viscous damping of $r$-mode oscillations in rapidly rotating compact
stars that contain quark matter.  We present results for the
color-flavor-locked (CFL) phase of dense quark matter, in which the
up, down and strange quarks are gapped, as well as the normal
(ungapped) quark phase. While the ungapped quark phase supports a 
temperature window $10^8$ K $\leq T\leq 5\times
10^9$ K where the $r$-mode is damped even for rapid rotation, the
$r$-mode in a rapidly rotating pure CFL star is not damped in the
temperature range $10^{10}$ K$\leq T\leq 10^{11}$ K.  Rotating hybrid
stars with quark matter cores display an instability window whose
width is determined by the amount of quark matter present, and they
can have large spin frequencies outside this window.  Except at high
temperatures $T\geq 10^{10}$~K, the presence of a quark phase allows
for larger critical frequencies and smaller spin-periods compared to
rotating neutron stars. If low-mass X-ray binaries contain a large
amount of ungapped or CFL quark matter, then our estimates of the 
$r$-mode instability suggest that there should be a population of
rapidly rotating binaries at $\nu\gtrsim 1000$ Hz which have not yet
been observed. 

\end{abstract}
 
\maketitle

\section{Introduction}
\label{sec_intro}

Neutron stars are astrophysical test-beds for the physics of
strongly-interacting matter in regimes of density that are not
presently accessible either in terrestrial experiments or by numerical
studies of lattice-regularized QCD. Therefore, connecting theoretical
advances in the knowledge of the equation of state (EoS) of dense
matter to astrophysical observations of neutron stars is essential.
The EoS, which plays an important role in determining a compact star's
properties and in its evolution with time, is constrained by
laboratory data only up to densities of about $2\rho_N$, where
$\rho_N\approx 2.6\times 10^{14}\ {\rm g~cm^{-3}}$ is the saturation
density of nuclear matter. At asymptotically high densities where
perturbative calculations in QCD are reliable,  quark matter is conjectured
to be in a color superconducting state commonly called the color-flavor-locked
(CFL) phase~\cite{Alford98}.  Meanwhile, models of low-energy QCD
at moderate baryon chemical potentials $\mu_B\gtrsim 1$ GeV in
a restricted temperature domain $T\leq T_c\sim 50$ MeV predict a
rich patchwork of color superconducting phases in the QCD phase
diagram~\cite{Steiner02,Schmitt,Shovkovy,Sharma,Neumann},  with
important consequences for the properties of self-bound quark stars or
neutron stars whose core density is large enough for deconfined quark
matter to exist there. We refer to the former as strange stars and
the latter as hybrid stars throughout this work.

The effects of color superconductivity on the EoS are small, though
potentially important for the surface structure of strange
stars~\cite{Lugones:2002va, JRS05,ARSS06}, or the nuclear-quark interface in hybrid
stars~\cite{ABPR05}. More interesting are its effects on transport
phenomena, as pairing among the up, down and strange quarks breaks the
$SU(3)$ global chiral symmetry spontaneously~\cite{Alford98}, 
leading to (pseudo-)
Nambu-Goldstone modes that determine the material response to external
perturbations. See the recent review Ref.~\cite{Alford:2007xm} for the
effective theory of these Nambu-Goldstone bosons. 
The pairing pattern that leaves the
maximal residual symmetry  in the three-flavor case is the  CFL
phase. In this phase, the up, down and strange quarks of all three
colors participate in Cooper  pairing and the order parameter reduces
the QCD symmetry $SU(3)_L\times SU(3)_R\times SU(3)_C$ to the diagonal
subgroup $SU(3)_{L+R+C}$ locking the color and flavor
symmetries. Apart from the $8$ pseudo-Nambu-Goldstone bosons
associated with the spontaneous breaking of chiral symmetry, other
light modes in this phase are the exactly massless Nambu-Goldstone
boson (the superfluid phonon), which is a long-wavelength fluctuation
of baryon number, and the massless photon, which is effectively a
combination of the vacuum photon and a gluon.

Studies of transport coefficients reveal that the thermal conductivity
and specific heat of the CFL phase at temperatures much smaller than
$T_c$, the critical temperature for pairing, is controlled by the
phonon and the photon, while the electrical conductivity is determined
by thermally excited electron-positron pairs~\cite{Shovkovy:2002kv}.
In the CFL phase the long-term cooling of the star is controlled by
neutrino emission from the scattering and weak decay of
phonons~\cite{JPS}. These novel transport phenomena imply  that
neutrino transport and neutron star cooling curves have the  potential
to distinguish between the numerous possible patterns of Cooper
pairing that might occur at stellar
densities~\cite{Jot,Blaschke1,Blaschke2,JRS06}.

Another set of transport coefficients that are modified by Cooper
pairing are the bulk and shear viscosities. Bulk viscosity can lead to
energy dissipation when the rate of chemical equilibration in matter
is comparable to the driving frequency of the volume perturbation.
Shear viscosity determines the relaxation of momentum components
perpendicular to the direction of fluid flow. Together, they
characterize the material response to compressional and shearing
forces that are typically present in rotating and pulsating compact
stars, and serve to damp out large-amplitude oscillatory modes of the
fluid. Moreover, since the fluid inside a star is self-gravitating,
these oscillations can couple to metric perturbations, causing the
star to emit gravitational waves. Indeed, the primary physical
interest in exploring viscosities of color superconducting quark
matter is to assess their role in damping out the so-called $r$-mode, a
particular oscillatory mode of the fluid that dissipates the star's
rotational energy by efficiently coupling the angular momentum of the
star to gravitational waves~\cite {Andy, Friedman, Morsink}. The
damping timescale, which depends sensitively on the viscosity, and
hence on the low-energy modes of the color superconductor determines
if the $r$-mode will be driven unstable. The unsuppressed growth of the
$r$-mode eventually saturates due to non-linear effects~\cite{Arras03}
since energy can be transferred from $r$-modes to ``inertial''
oscillation modes which do not couple to gravitational waves. However,
the saturation amplitude can still be large enough to emit
gravitational wave signals that can be detected directly by Advanced
LIGO~\cite{AK}.

The observational context of $r$-modes is that neutron stars do not
typically spin at rates near the maximum allowed frequency, the Kepler
frequency $\Omega_K$. $r$-modes offer a possible explanation of this
fact: in rotating neutron stars, these modes lose energy through
gravitational waves, which carry away angular momentum from the star
and act as braking radiation. We mention two distinct settings in which
$r$-modes can be instrumental in slowing the neutron star's rotation
speed.

(a) In the newly-formed isolated neutron star in the aftermath of a
core-collapse supernova: Whether $r$-modes are relevant for neutron
stars  at birth is difficult to determine because the number of
rapidly spinning pulsars that can be detected close to their birth is
a  negligible fraction of the total detectable sample. Limited by this
selection effect, most  inferences of rotation rates of neutron stars
rely on a model of time evolution of radio pulsars from birth
(neglecting any contribution from $r$-modes), with conflicting results
on the birth spin-periods, ranging from milliseconds~\cite{ARZ} to
hundreds of milliseconds~\cite{Kaspi}.  An independent method of
determining the initial spin period based on correlating a supernova's
observed X-ray luminosity (or upper limits thereof) with the
rotational energy loss of the neutron star suggests that a population
of millisecond pulsars at birth is ruled out~\cite{Stella}. On the
other hand, simulations of Type II supernovae including rotational effects 
predict relatively fast initial spin periods~\cite{Ott06}, which is 
difficult to reconcile with observed young and slowly rotating pulsars 
(like the Vela and Crab pulsars).  Thus, $r$-modes may play an important 
role in neutron stars at birth, and the impact of $r$-modes depends 
critically on the composition of the neutron star core. The critical 
temperature for color superconductivity being of the order of a 
proto-neutron star's temperature, it is important to understand the 
$r$-mode in the quark superfluid as it can determine the initial 
spin-down evolution of the strange star or hybrid star.

(b) In old neutron stars spun-up by accretion in binary systems:
Millisecond pulsars (typically, pulsars with spin-periods $\sim
10$~ms) are old and most of them are observed to be in low-mass X-ray
binary systems (LMXBs)~\cite{Lorimer}. The fast rotation is thought to
result from spin-up due to steady accretion from a low-mass companion
star. This picture has recently received confirmation through the
observation of pulsations in LMXBs. Spin frequencies in LMXBs, as
determined either from X-ray burst oscillations or from kHz
quasi-periodic oscillations, are typically between 300-640 Hz, well
below the maximum rotational rate (see Ref.~\cite{Lamb} for a recent
review). The upper limit on the rotation rate of LMXBs has been
recently revised by the detections of objects spinning at 716 Hz
~\cite{Hessels06} and 1122 Hz~\cite{Kaaret07}. However, the
statistical significance of the latter observation is not very
strong. One possibility is that accretion generates a mass quadrupole
moment which results in gravitational waves and this prevents the star
from rotating too quickly~\cite{Lars}. Upper limits on the generation
of gravitational waves from this mechanism have already been generated
by LIGO~\cite{Ben}. $r$-mode oscillations can be an alternative
explanation for the limiting frequency. The observation of these
fast-rotating objects provides an important constraint on the extent
to which $r$-mode oscillations can spin down neutron stars.  Since
inferred temperatures of an LMXB's interior are much less than $T_c$
for a quark superfluid, it is pertinent to ask how quark matter inside
an accreting neutron star affects the $r$-mode and the star's rotation
speed, assuming the quark core is not spun out of existence.

Motivated by these reasons, in this article, we investigate the effect
of  quark matter on $r$-modes in compact stars. In section~\ref{rmode},
we calculate the  $r$-mode frequency for perturbed rotating fluids that
obey chosen nuclear or quark matter EoS. In section~\ref{viscosity},
we review the viscosities of neutron matter, ungapped and gapped CFL
quark matter, which are inputs in a calculation of the damping
timescale of the $r$-mode. In section~\ref{damping}, we present results
for $r$-mode damping times in strange and hybrid stars, and discuss the
critical spin-frequency in section ~\ref{critical}. In
section~\ref{summary},  we remark on observational consequences for
such compact stars and present our conclusions.

\section{$r$-modes in hadronic and quark matter}
\label{rmode}

Various kinds of pulsation modes exist in neutron stars,
classified by the nature of the restoring force (see Ref.~\cite{Sterg} for
a review). For example, $p$-modes are high-frequency (few kHz)
pressure waves where fluid oscillations are largely radial, while
$g$-modes are low-frequency ($\sim$few 100 Hz) density waves driven by
gravity which tends to smooth out composition and thermal gradients, 
particularly in proto-neutron stars. The $r$-mode is
intimately linked to the rotational properties of the star, and the
restoring force here (in rotating stars) is the Coriolis force. The
modes are termed quasi-normal when they lose energy to gravitational
waves and may be classified by the parity of the fluid displacement
vector: polar (spheroidal) or axial (toroidal). $r$-modes are purely
toroidal only for the trivial case of a non-rotating star where they
have zero frequency. In a rotating star, the fluid displacement vector
corresponding to the $r$-mode acquires spheroidal components as well,
complicating the mode analysis. However, it is conventional to assume that
the modes in rotating stars remain toroidal. It is also conventional to 
apply the Cowling
approximation~\footnote{The Cowling approximation is equivalent to
neglecting  $\delta\phi_0$ in Eq.~(\ref{kappaeqn}). We include
$\delta\phi_0$,  however small, so we do not use this approximation.}
in which the back-reaction of the fluid perturbation on the metric,
and hence the gravitational potential, is ignored.  Previous
studies~\cite{Provost,Saio} have shown that including such effects
modifies the $r$-mode frequency approximately at the 5\% level.
Furthermore, if the perturbation is assumed to be isentropic, $\delta
P$ and $\delta\rho$ obey the same EoS as the unperturbed quantities:
$P$ (the pressure) and $\rho$ (the density). With these
approximations, the $r$-mode frequency in the co-rotating frame, to first
order in the rotation frequency of the star $\Omega$, is given 
by~\cite{Provost}
\begin{align}
\label{modefreq}
\omega_{rot} =\frac{2m\Omega}{l(l+1)} +\mathcal O(\Omega^3) .
\end{align}
 Since we are
interested in the instability to gravitational wave emission, we
restrict ourselves to the ``classical'' $r$-modes of Papaloizou and
Pringle~\cite{PP} for which $l=m$. An inertial observer measures a
$r$-mode frequency of
\begin{align}
\label{modefreq2}
\omega_r^{(0)} = \omega_{rot}-m\Omega= \[\frac{2}{l(l+1)}-1\] m\Omega
+\mathcal O(\Omega^3),
\end{align} from which it can be deduced that, for $l=m\geq 2$, a
counter-rotating mode in the rotating frame appears as co-rotating
with the star to a distant inertial observer. Thus, all $r$-modes with
$m\geq2$ are generically unstable to the emission of gravitational
radiation and the $m=2$ $r$-mode is the first to go unstable.  This is
the Chandrasekhar-Friedman-Schutz (CFS)
mechanism~\cite{Chandrasekhar70,Friedman78}. This instability is
active as long as its growth-time is much shorter than the damping time due
to the viscosity of stellar matter. Its effect is to slow the
rotation rate of a compact star in a short time span of about a
year~\cite{Morsink}. This possibly explains why only slowly rotating
pulsars are associated with supernova remnants. The $r$-mode
instability might not allow millisecond pulsars to be formed after a
supernova or an accretion induced collapse of a white dwarf; rather it
seems that millisecond pulsars can only be formed by the accretion
induced spin-up of old, cold neutron stars. However, in such cases,
the neutron matter inside the neutron star could be in a superfluid
state, and conclusions for $r$-mode damping in a superfluid phase are
strongly model-dependent~\cite{Mendell}.

Studies of $r$-mode oscillations provide a unique avenue to 
sample the star's density profile in addition to its  mass 
and radius. 
 $r$-mode oscillations distinguish between 
``pure'' neutron and hybrid or strange stars with the same mass and radius. 
To leading order in $\Omega$, 
the $r$-mode frequency has no dependence on the EoS, 
Eq.~(\ref{modefreq}). Including second-order rotational effects, one 
finds the following relation between the mode frequency $\omega_r$ in the 
inertial frame and $\Omega$~\cite{LMO} 
\begin{align}
\label{kappa2} \omega_r &=  \omega_{rot}-m\Omega\equiv\kappa\Omega-m\Omega, \\
\kappa & = \frac{2}{m+1}+\kappa_2\frac{\Omega^2}{\pi G\bar{\rho}_0}
+\mathcal O(\Omega^4), \nonumber
\end{align}
where $\kappa_2$ is obtained explicitly from Eq.~(\ref{kappaeqn}) below,
$\bar{\rho}_0$ is the average density of the unperturbed star and $G$
is Newton's constant. Ignoring general relativistic effects,
$\bar{\rho}_0$ is related to the Kepler frequency as
$\Omega_K=\frac{4}{9}\sqrt{2\pi G\bar{\rho}_0}$~\cite{Shapiro:1984}. $\kappa_2$ depends on
the density profile of the star, as well as density and
gravitational perturbations, which are related by the Poisson
equation for (Newtonian) gravity. We determine $\kappa_2$ from 
the second-order analysis of $r$-modes~\cite{LMO}
\begin{align}
\label{kappaeqn}
\kappa_2 \int_0^R  dr \left(\frac{r}{R}\right)^{2m+2} r 
\frac{d \rho}{ dr} =&{} \frac{6 m}{\left(m+1\right)^2} \int_0^R dr 
\rho_{22} \left(\frac{r}{R}\right)^{2m+2}\\ 
&{}+ \frac{8 \pi G \bar{\rho}_0 m}{ \left(m+1\right)^4} 
\int_0^R dr  r^2 \left(\frac{r}{R}\right)^{m+1}
\left[ \left(\frac{r}{R}\right)^{m+1}+\delta \Phi_0 \right]
\left(\frac{d \rho}{d h}\right)_0  ,\nonumber
\end{align}
where $0\leq r\leq R$ is the radial coordinate in the star,
$\rho_{22}(r)$ is the radial dependence of the non-isotropic
correction to the density of the rotating configuration to
lowest order in $\Omega$ and $\delta \Phi_0(r)$ describes the radial
dependence of the change in the gravitational potential due to the
$r$-mode oscillation, again to lowest order in $\Omega$. The quantity
$h$ is the barotropic enthalpy
\begin{align}
h(p)=\int_0^{p} \frac{dP^{\prime}}{\rho(P^{\prime})},  
\end{align}
 and $(d\rho/dh)_0$ denotes a derivative in the unperturbed star.
The function $\delta \Phi_0(r)$ is determined
through the solution of the differential equation 
\begin{align}
\label{delphi}
\frac{d^2\delta\Phi_0}{dr^2}+\frac{2}{r}\frac{\delta\Phi_0}{dr}+
\left[4\pi G\left(\frac{d\rho}{dh}\right)_0-\frac{(m+1)(m+2)}{r^2}\right]
\delta \Phi_0=-4\pi
G\left(\frac{d\rho}{dh}\right)_0\left(\frac{r}{R}\right)^{m+1},
\end{align}
whose derivation is detailed in Ref.~\cite{LMO}.
The function $\rho_{22}(r)$ is related to $\Phi_{22}(r)$, 
the change in the gravitational potential due to centrifugal deformation 
to lowest order in $\Omega$, 
as $\rho_{22}(r)=\left(\frac{d\rho}{dh}\right)_0[\Phi_{22}(r)-
\frac{1}{3}\pi G \bar{\rho}_0r^2]$, where $\Phi_{22}(r)$ obeys~\cite{LMO}
\begin{align}
\label{phi22}
\frac{1}{r^2}\frac{d}{dr}\left(r^2\frac{d\Phi_{22}}{dr}\right)-
\frac{6}{r^2}
\Phi_{22}+4\pi G\left(\frac{d\rho}{dh}\right)_0\Phi_{22}=\frac{4}{3}
\pi^2G^2r^2\bar{\rho}_0\left(\frac{d\rho}{dh}\right)_0 .
\end{align}
 Since the perturbations must fall to zero as 
$r\rightarrow\infty$, smoothness at the surface $r=R$ is enforced 
via logarithmic derivative matching 
\begin{align}
\left(\frac{d\delta\Phi_0}{dr}\right)_{r=R}=&-\left[\frac{1}{2}+
\sqrt{\frac{1}{4}+(m+1)(m+2)}\right]\frac{\delta\Phi_0|_{r=R}}{R},\\
\left(\frac{d\Phi_{22}}{dr}\right)_{r=R}=&-\frac{3\Phi_{22}|_{r=R}}{R}
. \nonumber 
\end{align}
  Along with the condition that the perturbations vanish at $r=0$,
Eqs.~(\ref{delphi}),(\ref{phi22}) fix $\rho_{22}(r)$ and
$\delta\Phi_0(r) $ throughout the star and subsequently $\kappa_2$
through Eq.~(\ref{kappaeqn}). We determine $\kappa_2$ and hence the
$r$-mode frequency for the $l=m=2$ mode since this is the lowest $l$
that can couple to gravitational waves. We consider four equations of state;
the first two are for hadronic matter, the next two are for quark matter.

1. The Akmal-Pandharipande-Ravenhall (APR) hadronic
EoS~\cite{Akmal98} whose microscopic input is based on the Argonne
$v_{18}$ nucleon-nucleon interaction~\cite{Wiringa} which is
calibrated to deuteron properties and vacuum nucleon-nucleon phase 
shifts for laboratory energies $E_{\rm lab}$ up to 350 MeV . 
The inclusion of the Urbana IX three-body force~\cite{Pudliner} 
and a relativistic boost term $\delta v_b$~\cite{Forest} successfully 
reproduces binding energies of several light nuclei.

2. To make contact with earlier works, we also employ a polytropic EoS
\begin{align}
P=K\rho^{1+1/n} ,
\end{align} where $K$ is a dimensionful constant for finite values of
the polytropic index $n$. The case  $n=0$  denotes incompressible
matter, which softens with increasing $n$. In Newtonian gravity,
stable configurations exist  only for $\Gamma=1+1/n>4/3$, i.e., for
$n<3$. We do not include  general relativistic corrections in the
structure or in the analysis of $r$-modes. Omitting these corrections
leads to a smaller mass and larger radius  for the compact star (see
values of radius $R$ in Table~\ref{timetable}).  The effect of general
relativistic corrections on the $r$-mode spectrum remains to be analyzed.

3. An MIT Bag model EoS for charge-neutral self-bound
ungapped quark matter, with Bag constant $B$ and current quark mass
$m_s$. The Bag constant is chosen such that 
the energy per baryon $E/A<930$ MeV at $P$=0. The
pressure is given by
\begin{align}
P=-\frac{3}{\pi^2}\sum_{i=u,d,s}\int_0^{k_{F_i}}dk~k^2
(\sqrt{k^2+m_i^2}-\mu_i)+\frac{\mu_e^4}{12\pi^2}-B .
\end{align}
For the parameters used in the calculation $m_u\sim m_d\ll m_s\ll
\mu_i$, we can write in perturbation 
\begin{align}
P= -B+\frac{\mu_e^4}{12\pi^2}
+\sum_{i=u,d,s}\frac{\mu_i^4}{4\pi^2}-\frac{3\mu_q^2m_s^2}{4\pi^2}+\mathcal
O(m_u^2, m_d^2, m_s^4), 
\end{align}
with quark chemical potential
\begin{align}
\mu_i=\mu_q- Q_i\mu_e,
\end{align}
where $Q_i$ is the electric charge and $\mu_q=(\mu_u+\mu_d+\mu_s)/3$
  is the average quark
chemical potential. Charge neutrality 
$\partial P/\partial\mu_e=0$ implies $\mu_e \approx m_s^2/(4\mu_q)$ and
we get
\begin{align}\label{pquark}
P&\approx \frac{3\mu_q^4}{4\pi^2}-B-\frac{3\mu_q^2 m_s^2}{4\pi^2},\\
n_q&=n_u+n_d+n_s=\frac{\partial P}{\partial\mu_q}\approx
\frac{3\mu_q^3}{\pi^2}-\frac{3\mu_q m_s^2}{2\pi^2}. \nonumber
\end{align}
The energy density $\epsilon = -P+n_q\mu_q$ yields
\begin{align}\label{equark}
\epsilon \approx \frac{9\mu_q^4}{4\pi^2}+B-\frac{3\mu_q^2
  m_s^2}{4\pi^2}. 
\end{align}
Eliminating $\mu_q$ between Eqs.~(\ref{pquark}) and (\ref{equark})
gives the EoS 
\begin{align}
\label{qapprox}
P\approx\frac{1}{3}(\epsilon-4B)-\frac{m_s^2}{3\pi}\sqrt{\epsilon-B},
\end{align}
which when compared to the exact EoS, is accurate to about 2\% for the 
entire density range of interest in a strange star made of ungapped 
quark matter.

 4. The CFL phase of quark matter~\cite{Alford98} with non-zero quark
masses. This phase is characterized by a common quark Fermi momenta
$k_F$ and Cooper pairing among all three flavors and colors of
quarks. This implies that the number densities  of all three quark
flavors are equal, which enforces electric charge neutrality even
without electrons~\cite{Rajagopal00,Steiner02}.  The pressure in the
CFL phase is given by~\cite{ARRW}
\begin{align}\label{Pcfl}
P_{CFL}=-\frac{3}{\pi^2}\sum_{i=u,d,s}\int_0^{k_F}dk~k^2
(\sqrt{k^2+m_i^2}-\mu_q)+\frac{3\Delta^2\mu_q^2}{\pi^2}-B,
\end{align} where small contributions from the phonon and
pseudo-Nambu-Goldstone bosons have been neglected since they have a
negligible effect in  determining the stellar structure. For realistic
quark chemical potentials $\mu_q$, the strange quark's current mass
$m_s\gg m_u\sim m_d$ can cause a mismatch in the quark flavor chemical
potentials, but as long as the mismatch is much smaller than the
superfluid gap $\Delta$, Cooper pairing with a common Fermi momentum
$k_F$ is still favored. The condition for a common Fermi momenta
$m_s^2< 4 \mu_q \Delta$~\cite{Schafer:1999pb} is satisfied for the
realistic parameters we use in the calculation. The average quark
chemical potential $\mu_q=(\mu_u+\mu_d+\mu_s)/3$ determines the total
quark number density
\begin{align}
n_q = n_u+n_s+n_d=\frac{\partial
  P_{CFL}}{\partial\mu_q}=\frac{3}{\pi^2}(k_F^3+2\Delta^2\mu_q), 
\end{align} 
which is three times the number density for each flavor. In
perturbation where $m_u\sim m_d\ll m_s\ll \mu_q$, we have
\begin{align}
k_F= \mu_q-\frac{m_s^2}{6\mu_q}+\mathcal O( m_u^2,m_d^2, m_s^4),
\end{align}
with pressure and energy density
\begin{align}
P_{CFL}&\approx\frac{3\mu^4}{4\pi^2}-B +\frac{3\mu^2}{4\pi^2}(4\Delta^2-m_s^2),\\
\epsilon& =-P_{CFL}+n_q\mu_q\approx \frac{9\mu^4}{4\pi^2}+B-\frac{3\mu^2
  m_s^2}{4\pi^2}\nonumber.
\end{align} 
Just as for ungapped quark matter, at leading order in the
quark mass expansion, we have
\begin{align}\label{cflapprox}
P_{CFL}\approx \frac{1}{3}(\epsilon-4
B)+\frac{4\Delta^2-m_s^2}{3\pi}\sqrt{\epsilon-B}. 
\end{align}  This approximate form based on the expansion in
$m_s^2/\mu_q^2$ works better than expected due to the accidental
combination $(4\Delta^2-m_s^2)/\mu_q^2$ which is numerically
small. $\mathcal O(m_s^4)$ corrections simply lead to a
renormalization of the bag constant~\cite{ABPR05} and can be ignored.
Indeed a comparison to the exact EoS verifies that the approximate
form Eq~(\ref{cflapprox}) is accurate to within $0.1\%$ for the entire
density range of interest in a strange star.

For hybrid stars, we employ the APR EoS for the hadronic phase and
either the MIT Bag or CFL EoS. Without modifications, a combination of 
the equations of state above does not generate a stable stellar
configuration which contains a significant amount of quark matter and which 
also gives a sufficiently large maximum mass when
general relativistic effects are included. For this reason, we make two
additional modifications. Firstly, we compute the
hybrid star profiles using general relativity, even though the expressions 
for $r$-mode properties are derived in the Newtonian limit. While
a more consistent approach is desirable, the present approach is not
unreasonable because it is the composition of the star, not the gross 
structural details, which determines the conclusions about the
stability of the $r$-mode.  Secondly, we add a term $-3 c \mu_q^4/(4 \pi^2)$ 
to the pressure, which characterizes the ${\cal O}(\alpha_s)$ perturbative 
QCD correction to the pressure of a non-interacting quark gas~
\cite{Fraga01,ABPR05} with $0<c<0.3$.

\begin{figure}[ht!]
\includegraphics[scale=0.6]{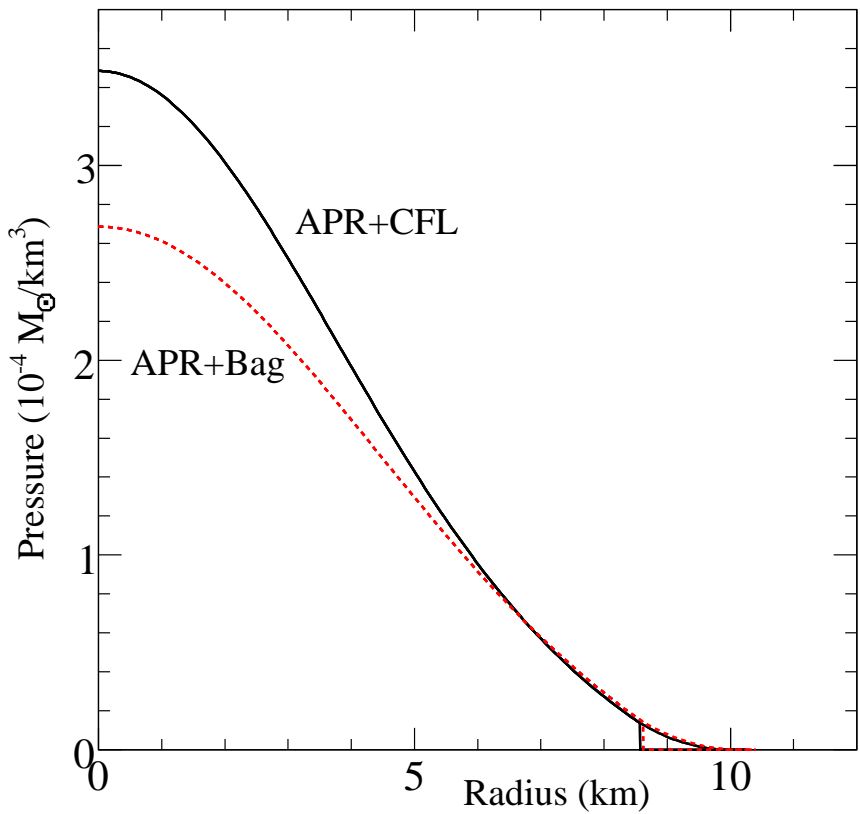}
\caption{The density profile of hybrid stars containing ungapped
(APR+Bag) and gapped (APR+CFL) quark matter that are considered in
this work. The transition densities were fixed at 
0.32 fm$^{-3}$. The vertical lines demarcate the homogeneous
quark and hadronic phases.}
\label{fig:profile}
\end{figure}

For hybrid stars, the bag constant and the parameter $c$ are chosen
such that transition density between hadronic and quark matter comes
out to be about twice the nuclear saturation density and the maximum
mass in general relativity is 1.7 
$M_{\odot}$. For ungapped quark matter, $B=75$ MeV/fm$^3$
and $c=0.083$, and for gapped quark matter, $B=106$ MeV/fm$^3$ and
$c=0.12$.~\footnote{CFL matter has larger pressure at a fixed density
  than ungapped matter because of the gap; consequently, the Bag
  constant value must increase to ensure that the phase transition
  occurs at the same density in both cases.} Unless strange quark
matter is absolutely stable at zero pressure, lower transition
densities are incommensurate with the nonobservation of deconfined quarks
in nuclei. Thus, the procedure outlined above effectively maximizes the 
amount of quark matter for a given hadronic equation of state in a hybrid 
star. Configurations with higher transition densities will have smaller 
quark matter cores, with results that lie between those for pure hadronic 
stars and hybrid stars with maximized quark content. Fig.~\ref{fig:profile} 
displays the typical density profile for hybrid stars containing ungapped 
or CFL quark matter for stars with mass $1.4 M_\odot$.

We assume that the surface tension between the quark and hadronic 
phases is large enough so that only one surface is created. 
An estimate of the critical surface tension and the microscopic nature
of the minimal CFL-nuclear interface is given in Ref.~\cite{ARRW}. The 
assumption of a sharp boundary is made in the interest of simplicity, 
in which case a Maxwell phase construction is sufficient to describe  
the surface, with pressures and chemical potentials 
being continuous  across the surface but density being discontinuous. 
A density discontinuity is clearly an idealization of a microscopically 
thin surface region where, for example, the number density will 
continuously transition from the larger value in the quark phase to 
the smaller value in the hadronic phase. This procedure is a good 
approximation if the surface tension is large enough that the energy 
gain provided by separating matter into a mixed phase is smaller than 
the surface energy cost.

For the equations of state discussed above, results for the
$r$-mode frequencies are listed in Table 1. We confirm earlier findings
for $r$-mode frequencies in neutron matter described by polytropic EoS,
viz., that the $r$-mode frequency for polytropes increases with
increasing softness of the EoS (higher polytropic index $n$). For
gapped or ungapped quark matter, the $r$-mode frequency is similar to
matter with a polytropic index $\approx 1.6$. The $r$-mode frequency
is robust to variations in the gap $0<\Delta/({\rm
  MeV})<100$, as long as other parameters are adjusted such that quark
matter remains self-bound.
\begin{table}[h!]
\begin{center}
\caption{$\kappa_2$ from Eq.~(\ref{kappaeqn}) and $m$=$2$ $r$-mode frequency
from Eq.~(\ref{kappa2}) for various equations of state. We
choose the central density such that the star's mass is
1.4$M_{\odot}$ in each case (unless otherwise specified) while the
rotation frequency is chosen to be $\Omega=\Omega_K$, where
$\Omega_K$ is EoS dependent. For quark matter in strange stars, we
choose $B=$80 MeV/fm$^3$, $m_s$=100 MeV and $\Delta=0$ MeV (ungapped) 
or $\Delta=$100 MeV (gapped). For hybrid stars, we choose
$B=75$~MeV/fm $^3$ and $c=0.083$ (ungapped) and $B=106$~MeV/fm$^3$
and $c=0.12$ (gapped). The fractional correction over the 
$0^{\rm th}$-order EoS-independent result $\omega_r^{(0)}$ given by
Eq.~(\ref{modefreq2}) is also displayed.}

\label{kappatable}

\begin{tabular}{lllll} \hline \noalign{\smallskip} $\hskip 0.5cm {\rm
EoS}$ & $\hskip 0.25cm\kappa_2$ &   $\hskip 0.4cm \omega_r^{-1} ({\rm
ms})$ &  $\hskip 0.5cm \left|\omega_r/\Omega\right|$ &   $\hskip
0.5cm\left|\frac{(\omega_r-\omega_r^{(0)})}{\omega_r^{(0)}}\right|$ \\
\noalign{\smallskip} \hline \hline \noalign{\smallskip}

$n=1 ~{\rm Polytrope}$  & $0.298$   &    \hskip 0.6cm $0.155$   
&    \hskip 0.6cm $1.215$   &    \hskip 1.0cm $0.088$\\

$n=2 ~{\rm Polytrope}$  & $0.148$ &  \hskip 0.6cm$0.147$ 
& \hskip 0.6cm$1.275$ & \hskip 1.0cm$0.044$\\

${\rm APR }$ 
  & $0.316$ &  \hskip 0.6cm$0.180$ 
& \hskip 0.6cm$1.208$ & \hskip 1.0cm$0.094$\\

${\rm APR}$~($M$=1.8 M$_{\odot}$)  & $0.266$ 
&  \hskip 0.6cm$0.161$ & \hskip 0.6cm$1.228$ & \hskip 1.0cm$0.079$\\

${\rm Bag}$  & $0.198$ &  \hskip 0.6cm$0.106$ 
& \hskip 0.6cm$1.255$ & \hskip 1.0cm$0.059$\\ 

${\rm CFL}$  & $0.195$ &  \hskip 0.6cm$0.099$
& \hskip 0.6cm$1.256$ & \hskip 1.0cm$0.058$ \\

${\rm APR+Bag}$   ($M$=1.7 M$_{\odot}$)  & $0.083$ &  \hskip 0.6cm$0.097$ 
& \hskip 0.6cm$1.301$ & \hskip 1.0cm$0.025$\\ 

${\rm APR+CFL}$  ($M$=1.7 M$_{\odot}$)   & $0.099$ &  \hskip 0.6cm$0.100$ 
& \hskip 0.6cm$1.294$ & \hskip 1.0cm$0.029$\\

\hline
\noalign{\smallskip} \hline
\end{tabular}
\end{center}
\end{table}

Since the $r$-mode is really a quasi-normal mode, the
mode frequency will generally acquire an imaginary part on account 
of the energy transmitted in gravitational waves and through dissipative 
forces in the fluid. To assess the $r$-mode instability and critical
rotation frequency of compact stars, a comparison of the 
viscous damping timescale with the mode-growth timescale due to 
gravitational wave emission is required. We proceed  
to a discussion of the role of viscosity in $r$-mode damping.

\section{Viscosity of hadronic and quark matter}
\label{viscosity}

The temperature-dependent bulk viscosity $\zeta$ and shear viscosity
 $\eta$ of the fluid suppress $r$-mode growth. For hadronic matter,
we use simple power-law fits derived in Ref.~\cite{Cutler87} which were
shown to faithfully reproduce results of microscopic calculations of
the viscosity~\cite{Flowers79}. For non-superfluid neutron matter, the
dominant contribution to the bulk viscosity comes from the modified
Urca process: $n+n\rightarrow n+p+e^-+ \bar{\nu}_e$
\begin{align}
\label{nbulkv}
\zeta_{\mathrm{mUrca}}=6\times 10^{25}\rho_{15}^2T_9^6\left
(\frac{\kappa\Omega}{\rm Hz}\right)^{-2} {\rm g/(cm\ s)} ,
\end{align}
 where the dimensionless quantities $\rho_{15}=\rho/(10^{15}~{\rm g/cm^3})$ 
and $T_9=T/(10^9~{\rm K})$. The dominant contribution to shear viscosity 
comes from $nn$ scattering and is given by
\begin{align}
\label{nshearv}
\eta_{nn}=2\times 10^{18}\rho_{15}^{9/4}T_9^{-2} {~\rm g/(cm\ s)} .
\end{align}

At low temperatures neutron matter is likely to be in a superfluid
state. At higher densities in the stellar interior, neutrons pair in
the $^3P_2$ state with $T_c\sim 10^8$ K~\cite{Schwenk04}  while they
pair in a $^1S_0$ state with
$T_c\sim 5\times 10^9$ K~\cite{TT93,Yakovlev01} in the outer
layers. At temperatures below the critical temperature
$T_c$ for superfluidity, the viscosity contributions from modified
Urca  processes are suppressed~\cite{Haensel:2001} so that   the
phonons and Nambu-Goldstone modes corresponding to spontaneous
breaking of rotational invariance (``angulons'') in the $^3P_2$
phase~\cite{Bedaque:2003wj} dominate the viscosity.  Ideally, the
viscosity estimates [Eqs.~(\ref{nbulkv}),~(\ref{nshearv})] for neutron
matter at $T\ll T_c$ should incorporate the effects of superfluidity.
We neglect these effects in the interest of simplicity, since the 
focus is on the quark phases. We note, however, that Lindblom \&
Mendell~\cite{Mendell} have shown that $\kappa_2$ (and hence the
$r$-mode frequency) for a superfluid neutron star described by the APR
EoS differs by less than 0.1\% from the non-superfluid case.
Furthermore, in computing the viscous damping timescales, these
authors have used electron scattering off magnetized vortices, termed
mutual friction~\cite{Alpar}, as the dominant source of bulk
viscosity, and electron-electron scattering as the dominant source of
shear viscosity. The results depend strongly on the model of
superfluidity in the core and introduce additional complications in
the analysis. We therefore postpone the practical issue of $r$-modes in 
superfluid neutron matter to future work.

For ungapped quark matter, the bulk viscosity is determined
mainly by the weak process $d+s\leftrightarrow u+s$~\cite{madsen},
while leptonic contributions can become important at high
temperature.~\footnote{At high temperatures $T\geq 10^{10}$~K, the
semi-leptonic processes $d(\mathrm{or}~s)\rightarrow
u+e^-+\bar{\nu}_e$ and  $u+e^-\rightarrow d(\mathrm{or}~s)+\nu_e$ can also
contribute significantly to the bulk viscosity~\cite{Basil}.  However,
for our parameter choices of $m_s=100$ MeV and $\omega_r$=1 kHz, this
contribution is negligible.} Since we are only examining stability
and not attempting to describe the full evolution of the $r$-mode, we
use the approximate expression for the bulk viscosity [Eq.~(16) of
Ref.~\cite{madsen}] which is appropriate for small oscillations of the
fluid and when $2\pi T\gg\delta\mu=\mu_s-\mu_d$ (this holds in the
range of temperatures of interest in this work). We have 
\begin{align}
\label{qbulkv} 
\zeta_{\,q}&=\frac{\alpha
T^2}{\omega^2+\beta T^4}\left[1-[1-{\rm
exp}(-\sqrt{\beta}T^2\tau)]\frac{2\sqrt{\beta}T^2/\tau}{\omega^2+\beta
    T^4}\right], \\
\alpha T^2&=\left(\frac{64}{45\pi^3}\right)G_F^2{\rm
sin}^2\theta_c{\rm cos}^2\theta_cm_s^4\mu_d^3T^2\nonumber\\
&=6.66\times
10^{20}\left(\frac{\mu_d}{\rm MeV}\right)^3\left(\frac{m_s}{\rm
MeV}\right)^4T_9^2 \quad {\rm g/(cm\ s^3)}, \nonumber\\ 
\beta T^4&=\frac{36}{\pi^2}\left(\frac{64}{45}\right)^2G_F^4{\rm
sin}^4\theta_c{\rm
cos}^4\theta_c\mu_d^6\left(1+\frac{m_s^2}{4\mu_d^2}\right)^2T^4\nonumber\\
&=3.57\times
10^{-8}\left(\frac{\mu_d}{{\rm
MeV}}\right)^6\left(1+\frac{m_s^2}{4\mu_d^2}\right)^2T_9^4\quad {\rm
s}^{-2},\nonumber
\end{align}
 where $\omega=\omega_r=2\pi/\tau$ is the frequency of the
perturbation, $G_F$ the Fermi constant and $\theta_c$ the Cabbibo
angle. In the bulk viscosity expression, the second term inside the
big square bracket is numerically negligible for the parameters of
the calculation. Further, this ``transient'' term drops out when
averaged over oscillation cycles leaving only the overall factor in
front, similar to the CFL bulk viscosity expression in
Eq.~(\ref{cflbulkv}).

The shear viscosity of ungapped quark matter is dominated by
quark-quark scattering in QCD. Neglecting small QED effects, we
have~\cite{Haensel89}
\begin{align}\label{qshearv} 
\eta_{\,q}\approx 6.99\times
10^{17}\left(\frac{0.1}{\alpha_s}\right)^{3/2}\left(\frac{\rho}{\rho_0}\right)^{5/3}
T_9^{-2}\quad
{\rm g/(cm\ s)} . 
\end{align}

In the superfluid CFL phase, the quarks are gapped. The massless
phonons and thermally excited light pseudo-Nambu-Goldstone bosons
determine the thermodynamic and hydrodynamic properties. Unlike the
mesons of the QCD vacuum, the mass ordering of the CFL
pseudo-Nambu-Goldstone bosons is ``reversed'' and the neutral kaon
$K^0$ is the lightest meson~\cite{Son:1999cm}. The contribution to
bulk viscosity from phonons alone has been calculated in
Ref.~\cite{Manuel:2007pz}. However, the dominant contribution to bulk
viscosity comes from flavor changing $K^0$ decay, viz., via
weak equilibrium processes $K^0\leftrightarrow \phi \phi$ and $\phi
K^0\leftrightarrow \phi$ involving the massless phonon
$\phi$~\cite{Alford:2007pj,Alf}. It should be noted that in 
contradistinction to normal matter, superfluid hydrodynamics is 
described by more than one bulk viscosity. Additional viscosities, 
commonly labeled $\zeta_1$ and $\zeta_3$, enter the dissipative terms 
in the stress-energy tensor, and they multiply factors of
$v_s-v_n$, the difference in the speed of the ``superfluid'' and
``normal'' component in the two-fluid model of
superfluidity~\cite{Khalatnikov,Landau}. The superfluid component is
non-viscous while the normal component consisting of the phonons and
other low energy excitations such as pseudo-Nambu-Goldstone modes is
viscous. Due to the large thermal conductivity of neutron and CFL
matter~\cite{Lattimer:1994,Shovkovy:2002kv}, we may approximate the
star to be isothermal. Consequently, the second sound associated with a
non-zero $v_s-v_n$ and temperature gradient would be damped, and
contributions from $\zeta_1$ and $\zeta_3$ can be
ignored in the leading order of this approximation. For the
calculations, we employ only the bulk viscosity
$\zeta_2$ (which becomes the usual bulk viscosity $\zeta$ in normal
matter) from neutral kaon $K^0$ decay~\cite{Alf}:
\begin{align}
\label{cflbulkv}
\zeta_{K^0}&=C\frac{\gamma_{\rm
eff}}{\omega^2+\gamma_{\rm eff}^2},\\
C&=\left(\frac{\partial
n_q}{\partial\mu_q}\right)^{-1}
\frac{\left(\bar{n}_K\frac{\partial
n_q}{\partial\mu_q}-\bar{n}_q\frac{\partial
n_K}{\partial\mu_q}\right)^2}{\frac{\partial
n_K}{\partial\delta\mu_K}\frac{\partial
n_q}{\partial\mu_q}-\left(\frac{\partial
n_K}{\partial\mu_q}\right)^2},\nonumber\\
\gamma_{\rm eff}&=\gamma_K \[ 1-\(\frac{\partial
  n_K}{\partial\mu_q}\)^2\(\frac{\partial n_q}{\partial\mu_q} 
\frac{\partial n_K}{\partial\delta\mu_K}\)^{-1}
\], \nonumber
\end{align}  where as before $\omega=\omega_r=2\pi/\tau$, and
$\gamma_{\rm eff}$ is  the effective kaon width~\cite{Alf}. The number
densities for quarks $n_q$ and  neutral kaons  $n_K$ are determined
from the thermodynamic pressure $P = P_{CFL}+P_K$ where the
contribution from kaons $K^0$ was added to the pressure $P_{CFL}$ from
Eq.~(\ref{Pcfl}). We choose parameters such that the effective kaon
chemical $\mu_K^{\rm eff}=(m_s^2-m_d^2)/(2\mu_q)$ is smaller than the
kaon mass $m_K$ ensuring that there is no kaon condensation. To compute the
bulk viscosity, one considers small departures from equilibrium
characterized by a non-zero chemical potential $\delta\mu_K$ and writes
\begin{align} P_K=&-\frac{T}{2\pi^2}\int_0^\infty dk k^2 \ln\[1-
\exp\(-\frac{E_K-\delta\mu_K}{T}\)\],\\ E_K=&-\mu_K^{\rm
eff}+\sqrt{\frac{p^2}{3}+m_K^2}\nonumber.
\end{align}
 The barred quantities and derivatives in Eq.~(\ref{cflbulkv})
 correspond to equilibrium values $\delta\mu_K=0$. The width
 $\gamma_{\rm eff}\approx\gamma_K$ is determined from the kaon decay  
 rate~\cite{Alf}
\begin{align}
\gamma_K&= \(\frac{\partial
  n_K}{\partial\delta\mu_K}\)^{-1}\frac{\Gamma_{\rm forward}(\delta\mu_K=0)}{T},\\
\Gamma_{\rm forward}&\approx\frac{|V_{ud} V_{us}|^2 G_F^2 f_\pi^2
  f_\phi^2}{9\sqrt{3}\pi} \(1+\frac{m_K^2}{{\mu_K^{\rm
  eff}}^2}\)\bar{p}^4
\frac{\exp\(\frac{\bar{p}}{\sqrt{3}T}\)}{\[\exp{\bar{p}/({\sqrt{3}T})}
-1\]^2} \nonumber,\\
\bar{p}&=\sqrt{3}\frac{m_K^2-{\mu_K^{\rm eff}}^2}{2\mu_K^{\rm
  eff}},\nonumber    
\end{align}
where $V_{ud}$, $V_{us}$ are the usual CKM
matrix elements. Perturbative estimates for the decay constants 
yield~\cite{Son:1999cm}
\begin{align}
f_K^2&=\frac{21-8\ln2}{18}\frac{\mu_q^2}{2\pi^2},\\
f_\phi^2&=\frac{3}{4}\frac{\mu_q^2}{2\pi^2}. \nonumber
\end{align}

In the bulk viscosity calculation we use $\Delta=100$ MeV, $m_u=5$
MeV, $m_d=7$ MeV, $m_s=100$ MeV and $B=80$ MeV/fm$^3$ to construct a
strange star in the pure CFL phase with mass $1.4~M_\odot$. 
A self-consistent determination
from  Eq.~(\ref{cflapprox}) gives $\mu_q\sim 300$ MeV and $\mu_K^{\rm
eff}\sim 17$ MeV. Unfortunately, at these moderate densities the kaon
mass $m_K$ is not accurately determined but it is expected to be
about $15-25$ MeV~\cite{Schafer:2002ty}.  If there is a region
in the star where $\delta m=m_K-\mu_K^{\rm eff} < 0$, then
modifications to the equation of state and viscosities due to kaon
condensation would be required~\cite{Alford:2008pb}, but to study a concrete 
case, we treat $m_K$ as a free parameter and choose
 $\delta m=m_K-\mu_K^{\rm eff}=1$ MeV, avoiding the issue of kaon 
condensation.  The CFL bulk viscosity drops rapidly with decreasing 
$T$ at higher values of $\delta m$.

\begin{figure}  \leavevmode
\includegraphics[height=7.5cm,width=7.5cm]{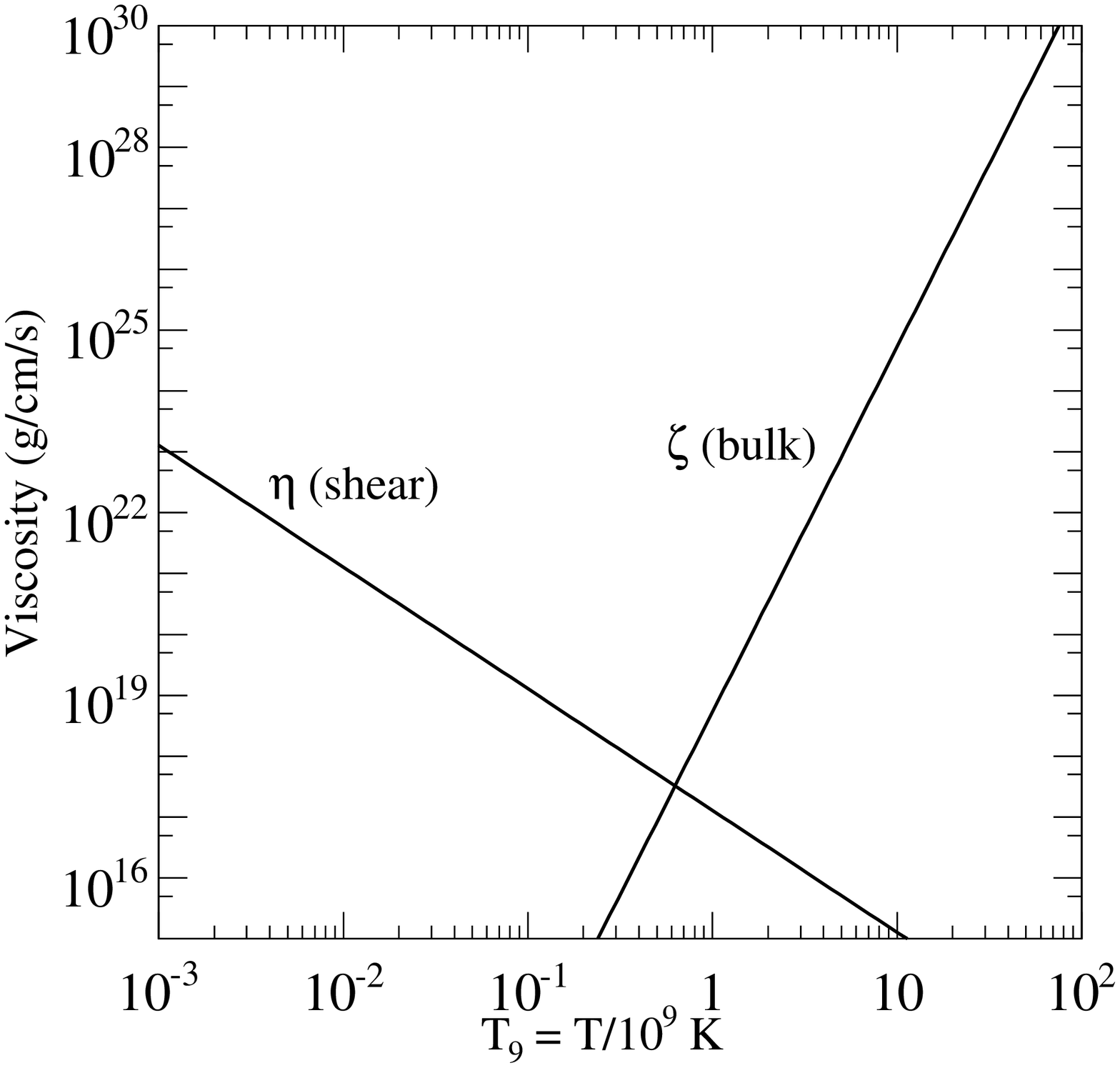}
\includegraphics[height=7.5cm,width=7.5cm]{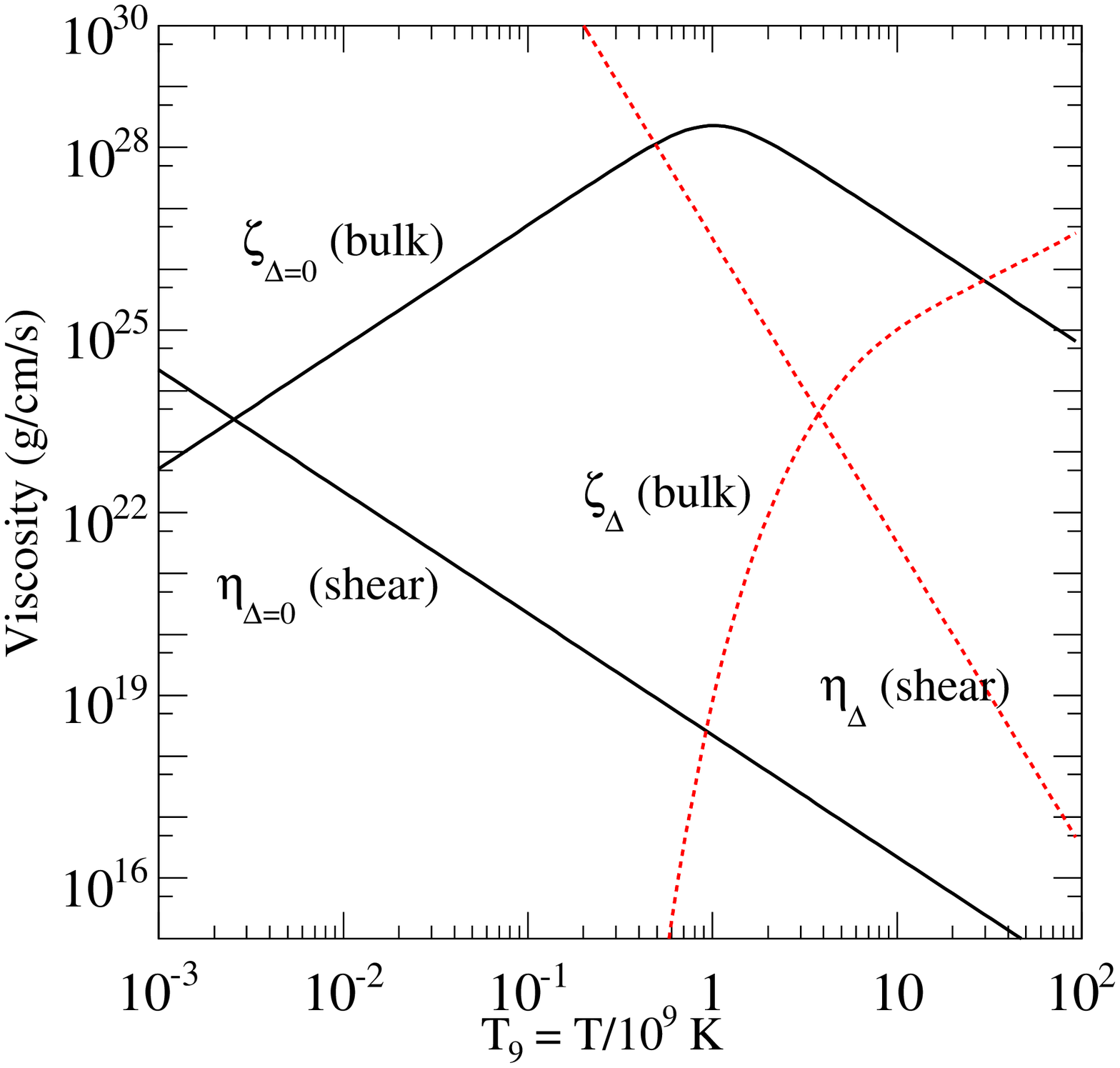}
\caption{The temperature dependence of the bulk and shear viscosity of
neutron matter from Eqs.~(\ref{nbulkv}), (\ref{nshearv}) (left panel) and 
that of
ungapped and gapped quark matter from Eqs.~(\ref{qbulkv}), (\ref{qshearv}) 
and 
Eqs.~(\ref{cflbulkv}), (\ref{cflshearv}) respectively (right panel). 
For neutron
matter, the energy density is chosen to be $1.5 \times
10^{-4}$ M$_{\odot}$/km$^3$ and $\kappa\Omega$ is fixed at 1000/s. For
quark matter, the quark chemical potential is chosen to be 310~MeV, $m_s$
=100 MeV with $\Delta$=100 MeV for CFL matter, and
$\tau = 2 \pi/\omega_r $ is fixed at 0.001~s.}
\label{fig:visc}
\end{figure}

The shear viscosity in the CFL phase has been calculated from phonon
scattering $\phi\phi\leftrightarrow\phi\phi$~\cite{CM}. The
contribution from  the medium modified photon was shown to be small
for temperatures where the phonon viscosity is of interest~\cite{CM},
therefore it is not included in the present calculation. A variational 
estimate of the linearized Boltzmann equation relevant for shear viscosity 
gives~\cite{CM}
\begin{align}\label{cflshearv}
 \eta_{\phi}=3.745\times 10^6 \(\frac{\mu_q}{\rm MeV}\)^8
T_9^{-5} {\rm\ g/(cm\ s)}.
\end{align} 
 The shear viscosity falls steeply with increasing
temperature and consequently  has a small effect on the viscous
damping of the $r$-mode at high temperature, see Fig.~\ref{fig:visc}
and  Eqs.~(\ref{InvShear}),~(\ref{criteqn}). Although the shear
viscosity is   undoubtedly more important at low temperatures, the
contribution from the phonon becomes irrelevant at very low
temperatures. This is because being a Nambu-Goldstone mode, the phonon
is derivatively coupled where for typical interaction terms
$i\partial_0\phi =E_\phi\phi\sim T\phi$ for thermally exited phonons.   
At very low temperatures $T\ll \mu_q$, then, the phonon
scattering cross section is suppressed and the mean free path
$\lambda_{\phi}$ increases. The hydrodynamic expression for the shear
viscosity is only relevant when $\lambda_{\phi}$ is much smaller than
the star radius $R$, otherwise the phonon travels through the star 
without collisions responsible for shearing the fluid flow.  
An order of magnitude
estimate of the relevant phonon mean free path can be obtained from
the kinetic theory relation $\eta_\phi\sim n_\phi p\lambda_{\phi}$,
where $n_\phi$ is the phonon number density and $p\sim 2.7 T/v$ is the
thermally averaged phonon momentum with phonon speed
$v=1/\sqrt{3}$. This length scale $\lambda_\phi$ quantify the distance
over which the photon must travel to generate shear viscosity due to
collision.  
We
estimate
\begin{align}
\lambda_\phi\sim \frac{\eta_\phi}{p n_\phi}\approx 2.1\times
10^9\(\frac{\mu_q}{300\ {\rm MeV}}\)^8 T_9^{-9}\ {\rm km}. 
\end{align} For typical quark chemical potential $\mu_q\sim 300$ MeV,
the phonon mean free path $\lambda_\phi$ will exceed the stellar radius at
temperatures $T\leq 10^{10}$ K. The calculated damping times for the CFL
phase are valid only above this temperature. For $T<10^{10}$~K, 
contributions other than from phonons are expected to be
relevant for shear viscosity but these have yet to be estimated. For
this preliminary work on the CFL $r$-modes, we have included only the
phonon contribution.

For hybrid stars, we ignore possible bulk viscosity due to the
quark-hadron  interface~\cite{Zheng06}, and simply utilize the
viscosity expressions  above for each phase. This approximation
implies that the viscosities for hybrid stars may be
underestimated compared to a realistic treatment of the
interface/crust.

\section{viscous damping of $r$-modes}
\label{damping}

The energy of the $r$-mode is dissipated according to~\cite{LOM}:
\begin{align}
\label{rmodeloss}
\frac{dE}{dt}=-(\omega_{rot}-m\Omega)^{2m+1}\omega_{rot}|\delta
J_{mm}|^2
-\int d^3r (2\eta\delta\sigma^{ab}\delta\sigma_{ab}+
\zeta\delta\sigma\delta\sigma) , 
\end{align}
where $\delta\sigma=\nabla_a\delta v^a$ is the volume
expansion due to the $r$-mode and $\delta\sigma_{ab}=\frac{1}{2}
(\nabla_a\delta v_b+\nabla_b\delta
v_a-\frac{2}{3}\delta_{ab}\nabla_c\delta v^c)$ is
the shear tensor. The first term is the energy radiated in
gravitational waves to lowest order in $\Omega$ with $\delta J_{mm}$ being
the current multipole 
\begin{align}  
\delta J_{mm}\propto
R^2\Omega\int_0^R dr \rho \left(\frac{r^2}{R}\right)^{(m+1)} . 
\end{align}
For $m\geq 2$, $\omega_{rot}<m\Omega$, 
so that the $r$-mode energy
grows with gravitational wave emission, triggering the instability.
The timescale $\tau$ associated with growth or dissipation is given by
\begin{align}
\label{damptime}
\frac{1}{\tau_i}=-\frac{1}{2 E}\left(\frac{dE}{dt}\right)_i .
\end{align}
The energy $E$ of the $r$-mode is given by
\begin{align}
\label{etotal}
E = \frac{\pi}{2 m} \left(m+1\right)^3 \left(2 m+1\right)!
R^4 \Omega^2 \int_0^R dr \rho \(\frac{r}{R}\)^{2m+2} ,
\end{align}
 where we have dropped a proportionality constant that determines the
amplitude of the $r$-mode, since it cancels in the evaluation of the 
damping timescale. Explicitly, the gravitational radiation time scale 
is~\cite{LOM}
\begin{align}
\frac{1}{\tau_{GW}} = - \frac{32 \pi G \Omega^{2m+2}}{c^{2m+3}} 
\frac{\left(m-1\right)^{2m}}{\left[\left(2m+1\right)!!\right]^2} 
\left(\frac{m+2}{m+1}\right)^{2m+2} \int_0^R dr \rho r^{2m+2} ,
\end{align}
 while the shear viscosity time scale is~\cite{LOM}
\begin{align}\label{InvShear}
\frac{1}{\tau_{\eta}} = \frac{(m-1)(2 m+1)}{\int_0^R dr \rho r^{2m+2}}
\int_0^R dr \eta r^{2m} .
\end{align}

 We compute the bulk viscosity timescales using a simplified expression
for the volume expansion, which is based on approximating
the Lagrangian perturbation of the fluid by an Eulerian one
~\cite{LOM}
\begin{align}
\label{euler}
\delta \sigma = -i \kappa \Omega \frac{\delta \rho}{\rho} .
\end{align}
As shown in Ref.~\cite{LMO}, while this approximation is only good
within an order of magnitude for the bulk viscosity damping timescale,
the critical angular velocity, which can be constrained by observations, 
is hardly modified because the bulk viscosity has a steep temperature 
dependence. Thus, the critical frequency curves presented in this work 
are accurate, even though the bulk viscosity timescales are only approximate.

Carrying out angular integrations for the bulk viscosity part of
Eq.~(\ref{rmodeloss}) and using Eq.~(\ref{euler}), one obtains
\begin{align}
\label{dedtbulk} \left(\frac{d \tilde{E}}{d t}\right)_{\zeta}  = - 2
\left[\frac{(2m+1)!}{2 m+3}\right] \int dr (2 \pi r^2) \zeta \frac{4
R^4 \Omega^6} {(1+m)^2 \rho_0^2} \left(\frac{d \rho}{d
h}\right)_{0}^2 \left[ \left(\frac{r}{R}\right)^{m+1}+ \delta
\Phi_0(r)\right]^2  ,
\end{align}
 where $\rho_0(r)$ is the density of the unperturbed star. 
The bulk viscosity damping timescale follows from
Eqs.~(\ref{damptime}), (\ref{etotal}) and (\ref{dedtbulk}).

\begin{figure}[ht!]
\leavevmode
\includegraphics[scale=0.4]{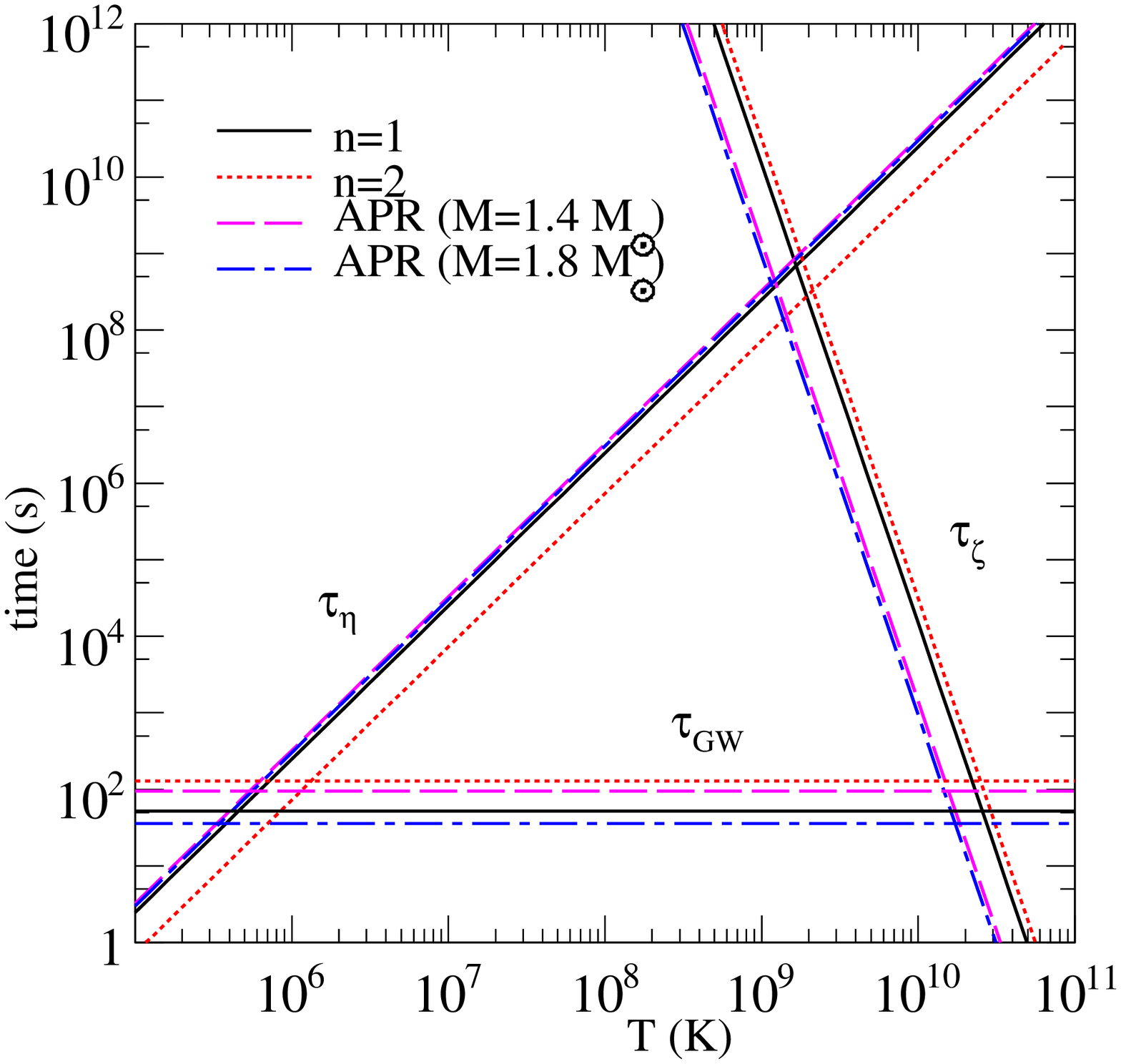}
\includegraphics[scale=0.4]{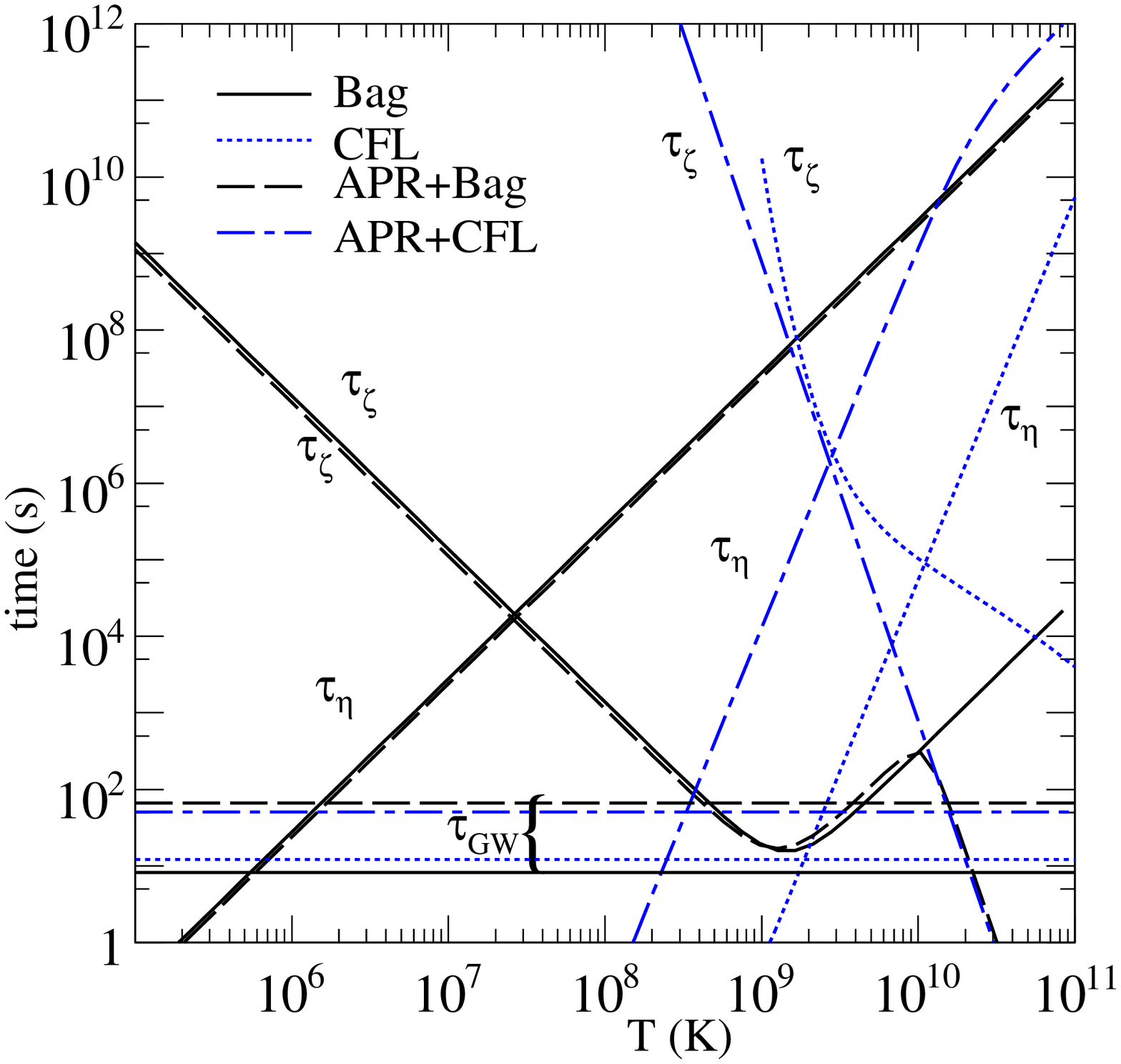}
\caption{The temperature dependence of the bulk and shear viscosity
damping timescales for neutron matter (left panel) and for strange and
hybrid star matter (right panel). The rotation rate is the Kepler frequency 
$\Omega_K$.}
\label{fig:damptime}
\end{figure}

For a fixed temperature $T$=$10^9$ K, and for stars rotating at the Kepler
frequency $\Omega=\Omega_K$, all three damping timescales are
displayed in Table~\ref{timetable}. The timescales are also
plotted as a function of temperature in Fig~\ref{fig:damptime}. We
observe from Fig.~\ref{fig:damptime} that in non-superfluid neutron
matter, shear viscosity damps the $r$-mode  rapidly only at very low
temperatures $T<10^{6}$ K, while bulk viscosity  damps it  only at
very high temperatures $T>10^{10}$ K.  In normal  quark matter, the damping
timescale due to bulk viscosity has a minimum around $T\sim 10^9$ K,
corresponding to a maximum in the bulk viscosity. This behavior leads
to a high-temperature stability window. In CFL quark matter, the
damping  timescale due to shear viscosity at low temperature is
large. However, the phonon contribution considered here is not
reliable below about $10^{10}$~K as mentioned earlier.    
Other contributions, such as from light-by-light scattering~\cite{CM}, could
become important in this regime and damp out the $r$-mode. 
Keeping this caveat in mind, and noting that the damping timescale 
due to bulk viscosity is still large except at 
high temperatures $T>10^{11}$~K, the $r$-mode in CFL matter is unstable
only in a small window around high temperatures $T\sim 5\times 10^{10}$~K, 
in marked contrast to either ungapped quark matter or normal neutron matter.
It would be interesting to see how the width of this window changes upon 
inclusion of relevant shear viscosities at temperatures $T<10^{10}$~K.

\begin{table}[ht!]
\begin{center}
\caption{Damping times for various equations of state. The central
energy density $\bar{\rho}_c$ in each case is chosen to yield a 1.4 
$\mathrm{M}_{\odot}$
star (unless otherwise explicitly stated)
and the temperature is fixed at $T$=$10^9$K. $\tau_f$ is the critical or
cumulative $r$-mode damping timescale given by 
$\tau_{\rm f}=(1/\tau_{\zeta}+1/\tau_{\eta}+1/\tau_{GW})^{-1}$. The
star is taken to be rotating at the Kepler frequency $\Omega_K$.
Energy densities are given in units of $10^{-4}
~\mathrm{M}_{\odot}/\mathrm{km}^3$.}
\label{timetable}
\begin{ruledtabular}
\begin{tabular}{lccccccc}
EoS  & 
$\bar{\rho}_c$ & $R$(km) &
$\Omega_K$(kHz) & $\tau_{\mathrm{\zeta}}$(s) & $\tau_{\mathrm{\eta}}$(s)
& $\tau_{\mathrm{GW}}$(s) & $\tau_{\mathrm{f}}$(s) \\
\hline
$n=1 ~{\rm Polytrope}$   & 5.63 &
12.50 & 5.31 & 2.47 $\times 10^{10}$ & 2.48 $\times 10^{8}$ &
$-$5.22 $\times 10^{1}$ & $-$5.22 $\times 10^{1}$ \\
$n=2 ~{\rm Polytrope}$  & 19.5 &
12.50 & 5.31 & 6.20 $\times 10^{10}$ & 7.28 $\times 10^{7}$ &
$-$1.29 $\times 10^{2}$ & $-$1.29 $\times 10^{2}$ \\
APR  & 3.60 &
13.74 & 4.61 & 2.48 $\times 10^{9}$ & 3.25 $\times 10^{8}$ &
$-$9.54 $\times 10^{1}$ & $-$9.54 $\times 10^{1}$ \\
APR ($M=1.8~\mathrm{M}_{\odot}$)  & 3.97 &
14.06 & 5.05 & 1.79 $\times 10^{9}$ & 3.03 $\times 10^{8}$ &
$-$3.58 $\times 10^{1}$ & $-$3.58 $\times 10^{1}$ \\
Bag & 4.30 &
9.89 & 7.54 & 7.33 $\times 10^{1}$ & 2.78 $\times 10^{7}$ &
$-$8.28 $\times 10^{0}$ & $-$9.34 $\times 10^{0}$ \\
CFL & 3.23 &
10.71 & 6.70 & 1.74 $\times 10^{10}$ & 5.38 $\times 10^{-1}$ &
$-$7.48 $\times 10^{-1}$ & 1.92 $\times 10^{0}$ \\
APR+Bag  ($M$=1.7 M$_{\odot}$) & 13.5 &
10.25 & 7.93 & 1.04 $\times 10^{2}$ & 2.38 $\times 10^{7}$ &
$-$7.96 $\times 10^{0}$ & $-$8.62 $\times 10^{0}$ \\
APR+CFL  ($M$=1.7 M$_{\odot}$)  & 10.6 &
10.40 & 7.69 & 1.72 $\times 10^{9}$ & 1.29 $\times 10^{4}$ &
$-$8.66 $\times 10^{0}$ & $-$8.67 $\times 10^{0}$
\end{tabular}
\end{ruledtabular}
\end{center}
\end{table}

\section{Critical rotation frequencies} 
\label{critical}

The critical rotation frequency $\Omega_c=1/\tau_{\rm c}$ of
neutron/strange/hybrid stars can be determined by the criterion that
at this frequency, the fraction of energy dissipated per unit time
exactly cancels the fraction of energy fed into the $r$-mode by
gravitational wave emission:
\begin{align}
\label{criteqn} \frac{1}{\tau_f}\Big|_{\Omega_c}=\left[\frac{1}{\tau_{\zeta}}+ \frac{1}{\tau_{\eta}}+\frac{1}{\tau_{\rm GW}}\right]\Big|_{\Omega_c}=0 . 
\end{align}

\begin{figure}
\bce
\leavevmode
\includegraphics[scale=0.4]{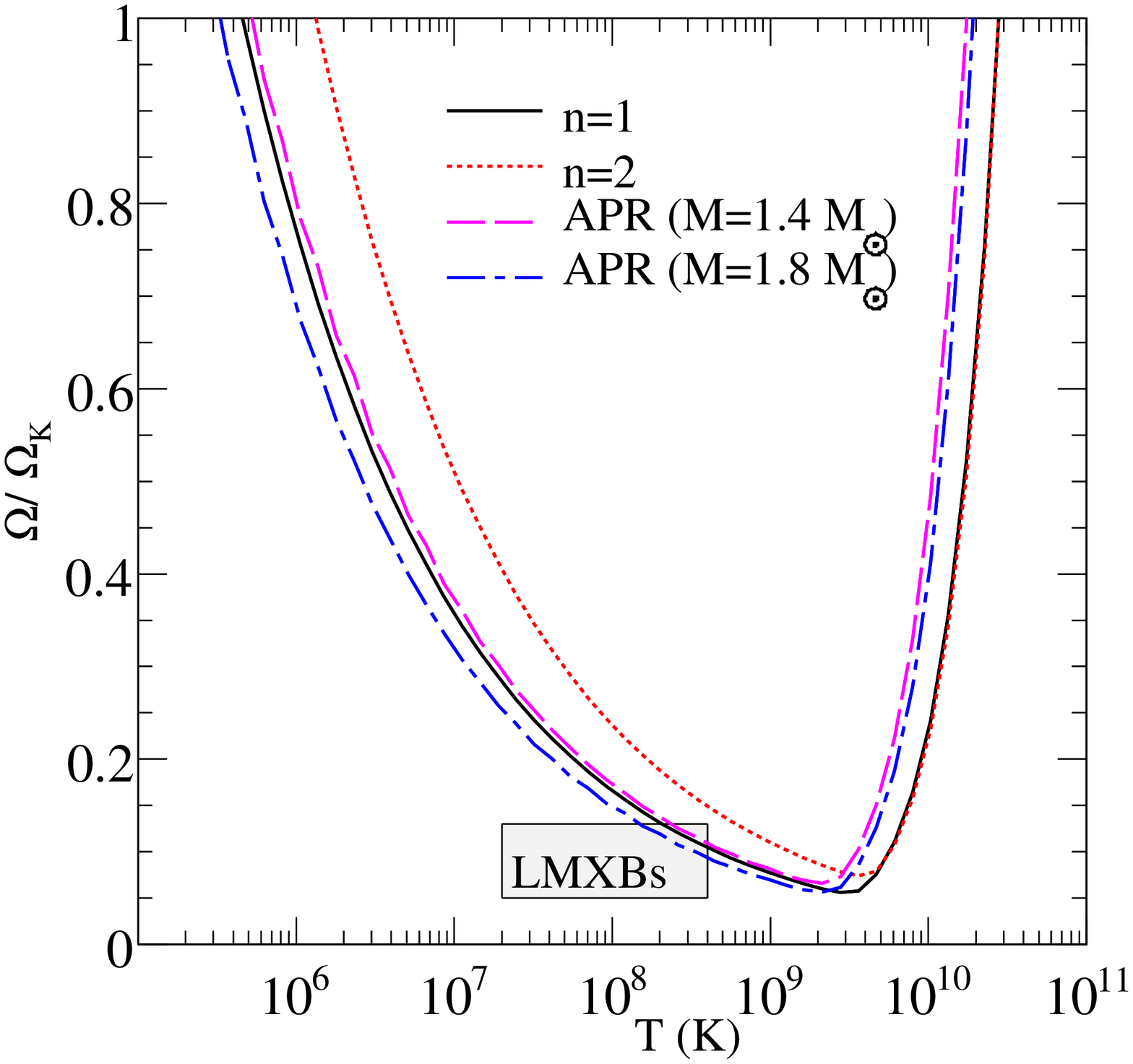}
\includegraphics[scale=0.4]{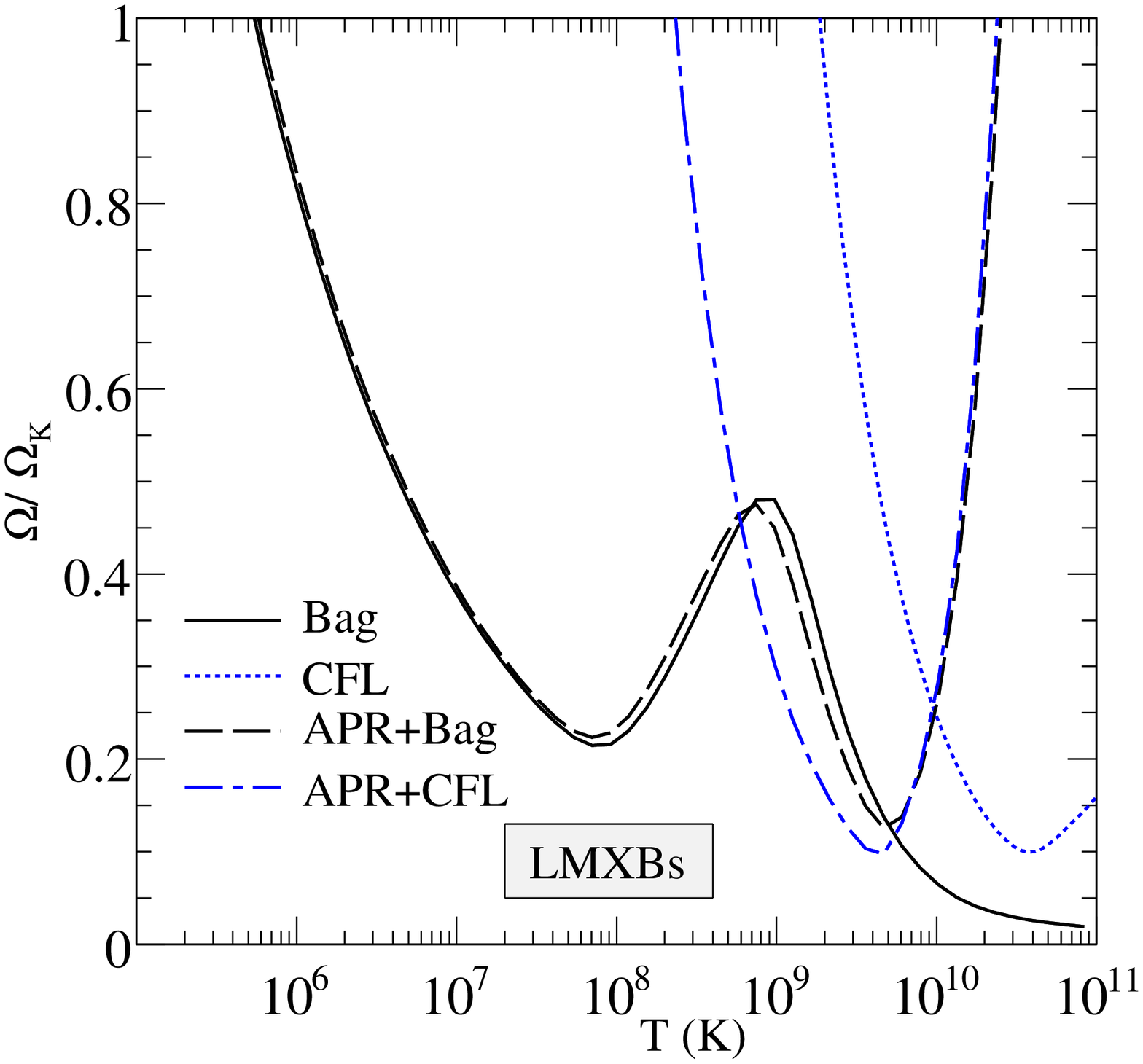}
\caption{The critical frequency $\Omega_{\rm c}$ in units of $\Omega_K$ 
 as a function of
 temperature for neutron stars, strange stars and hybrid stars. For quark
  matter, we choose $m_s$=100 MeV and $\Delta$=100 MeV. The bag
  constant $B$=80~MeV/fm$^3$ for strange stars, while $B$=110 (150)
  MeV/fm$^3$ for hybrid stars with ungapped (CFL) quark matter. The
  box represents typical temperatures 
  ($2 \times 10^{7}$-$3 \times 10^8$ K) and rotation
frequencies (300-700~Hz) of the majority of observed LMXBs assuming
 $\Omega_K=5500$ Hz.}
\label{fig:critfreq}
\ece
\end{figure}

Stable rotation frequencies at any temperature will be limited by the
smaller of the critical frequency or the Kepler limit
$\Omega/\Omega_K$=1.  As depicted in Fig.~\ref{fig:critfreq}, the
region above the temperature-dependent $\Omega_c$ curve is unstable to
$r$-mode oscillations and the star, if it enters this region, will be
spun down rapidly to $\Omega<\Omega_c$.

Normal neutron stars are generally unstable to $r$-modes except at very
high or very low temperatures, where the bulk and shear viscosities
respectively are effective at damping $r$-mode  oscillations. The width
of the instability window is not very sensitive to stellar mass, but is
noticeably smaller for a softer equation of state (larger polytropic
index). The box marked ``LMXBs'' is representative of typical LMXB
core temperatures~\cite{Brown02} $2 \times 10^7$ to $3 \times 10^8$ K,
with rotation frequencies between 300 and 700 Hz.  Strange stars made
of ungapped quark matter are more stable than their neutron star
counterparts in a window of temperatures $10^8$ K to $5 \times 10^9$~K
where the bulk viscosity damping timescale in quark matter is quite
small. Strange stars composed of CFL matter are   unstable to $r$-mode
oscillations in the range $10^{10}$ K$\leq T\leq 10^{11}$ K. Hybrid
stars with ungapped quark matter  are much like ungapped strange
stars, except that the bulk viscosity from the outer shell of hadrons
damps the $r$-mode oscillations at higher temperatures. $r$-mode
oscillations in hybrid stars with gapped quark matter are unstable in
the range $10^9$ K $\leq T\leq 10^{10}$ K.  The results for hybrid
stars are much like the corresponding strange star model because we
have maximized the amount of quark matter by fixing the deconfinement
transition density to be small. If this transition density is larger,
the critical curves for hybrid stars look more similar to those in the
left panel of Fig.~\ref{fig:critfreq}.

\section{Conclusions}
\label{summary}

We have calculated the frequency and viscous damping timescale of the
$m=2$ $r$-mode for compact stars made of hadronic matter, quark matter,
and for hybrid stars. To our knowledge, this is the first attempt at
studying $r$-mode characteristics for pure or hybrid stars with color
superconducting quark matter,  particularly the CFL phase. In absolute
terms, the $r$-mode frequency is about 50\% larger for strange stars in
comparison to neutron stars of the same mass with a polytropic EoS,
assuming both are rotating at their respective Kepler
frequency. The ratio of the $r$-mode frequency to the star's rotation
frequency is larger for strange stars and hybrid stars; this is due to
the softer equation of state.  It is also evident that EoS effects on
the mode frequency are smaller for softer equations of state.

The dominant bulk and shear viscosities of the normal hadronic
phase, the ungapped quark phase as well as the gapped CFL quark
phase, which was derived recently, are used to estimate viscous
damping timescales for homogeneous and hybrid stars. The current results
confirm established results for neutron stars; viz. that since neutron
stars are stable against the $r$-mode instability only at very high
($T\gtrsim$ 10$^{10}$ K) or very low temperatures ($T\lesssim$
10$^{6}$ K), they would be spun down rapidly by the $r$-mode
instability shortly after their birth at MeV temperatures, when large
neutrino losses cause rapid cooling. The predicted critical frequency
for LMXBs with hadronic neutron stars is close to  or inside the
region given by observations and LMXBs. A more careful analysis
taking the effect of superfluidity in neutron matter into account
would provide a clearer picture on the presence of $r$-modes assuming 
no quark matter is present.

Strange stars with non-superfluid quark matter
display a stability window between $10^{8}$~K~$<T<5\times 10^{9}$~K
where they can spin at a substantial fraction of the Kepler frequency.
If LMXBs contain strange stars, then on statistical grounds, we should
expect to observe some of them spinning close to their Kepler
frequency, unless some mechanism not connected to $r$-modes is
responsible for limiting their frequency or the quark phase is more
complicated than considered here (ungapped or CFL). If newly-born
superconducting strange stars occur in nature, they may still spin
down rapidly at birth, provided the cooling timescale is slower than
the spin-down timescale. A consistent analysis of spin-down and
cooling is currently under investigation and will be reported in
separate work. Hybrid stars tend to be more resilient against
the $r$-mode instability, except for a temperature range
which is dependent on the quark EoS. This effect is not surprising: in
hybrid stars, $r$-modes which would normally be undamped in hadronic
matter are damped by the quark core and vice versa. We conclude that
hybrid proto-neutron stars are not likely to be damped by $r$-modes at
all. This work also has important implications for LMXBs if they are
hybrid stars containing a significant amount of deconfined quark matter 
in the ungapped or CFL phase, or if LMXBs are strange stars with a thin
accreted hadronic crust. Based on the estimates of viscous damping 
timescales, the presence of quark matter in LMXBs implies that either 
they must contain rapidly rotating stars that have not yet been observed, 
or have a more complicated phase structure. A similar explanation regarding 
the LMXB rotating at 1122 Hz was put forth recently~\cite{Parenti07}.

To make the results more concrete, future work will have to address
several assumptions more closely. We assumed that the modified
Urca process provides the dominant contribution to bulk viscosity
while  neutron-neutron scattering determines the shear viscosity. In
practice, in the core of neutron stars, for $T\leq T_c\sim 10^8$ K,
neutron matter is  likely to be in a triplet-paired superfluid
state. This will suppress the  modified Urca rate~\cite{Haensel:2001},
rendering it insignificant at $T\ll T_c$.  Singlet pairing among
neutrons and/or protons has a larger critical temperature $T_c\sim
5\times 10^{9}$ K~\cite{Yakovlev01} and can occur in the  crust with
similar consequences. Under these conditions, it is more realistic
to include alternate sources of viscosity along the lines
of Ref.~\cite{Alpar}. It would also be useful to explicitly check 
the 
contributions from  the additional viscosities $\zeta_1$ and
$\zeta_3$ that are inherent in a 2-fluid model of superfluidity.

We also assume that in the CFL phase, the shear viscosity is from
phonons alone, whereas below $T\sim 10^{10}$ K, the damping
timescales are not reliable as the phonon mean free path exceeds the
stellar size. At these low temperatures, other contributions such as
from photons would be more relevant and need to be explicitly computed
and included for an accurate treatment of $r$-modes in CFL matter.

In the current work, 
we considered a simplified  model for a neutron/strange
star in this work, ignoring realistic features  such as a crustal
layer, rubbing friction at the quark-hadron interface and possibly the
role of other contributions to the bulk and shear viscosity at very
high or very low temperatures. Neutron star crusts composed of
hadronic matter typically damp $r$-mode oscillations further. Early work
in Ref.~\cite{Bildsten00} suggested that damping timescales were
decreased by as much as a factor of $10^5-10^7$ because of the
additional  viscosity provided by the neutron star crust. This
conclusion has been softened in some later works: Ref.~\cite{Levin01}
concluded that the associated decrease in the damping timescale was a
factor of $10^2-10^3$ smaller, and Ref.~\cite{Glampedakis06} has
pointed out that the damping timescale is sensitive to the crustal
composition. The final result may well be that $r$-modes are not as
important a mechanism as believed for  spinning down fast-rotating
compact objects. Ref.~\cite{Lin07} has also stressed the importance of
updating the multi-component nature of equation of state in order to
fully assess the effects of superfluidity. However, this does not
change the central conclusion, viz.,  if $r$-modes do play a role in
spinning down neutron stars, the present work demonstrates that this effect is
much less efficient in neutron stars containing ungapped or completely gapped
quark matter.

\section*{Acknowledgments}

The authors thank Mark Alford, 
Edward Brown, Cristina Manuel, Thomas Sch\"{a}fer
and  Andreas Schmitt for useful discussions related to this work.  PJ
acknowledges support from 
the Department of Atomic Energy of the
Government of India, the Argonne National Laboratory as a visiting
scientist under US Department of Energy, Office of Nuclear Physics,
contract no. DE-AC02-06CH11357, and the Ohio University Office of
Research. 
GR is supported in part by the US
Department of Energy grant DE-FG02-03ER41260. AWS is supported by the
Joint Institute for Nuclear Astrophysics at MSU under NSF-PFC grant
PHY 02-16783 and by NASA under grant NNX08AG76G.


\begin{thebibliography}{10}

\bibitem{Alford98}
M.~G. Alford, K.~Rajagopal and F.~Wilczek,
\newblock Nucl. Phys. {\bf B537}, 443 (1999).

\bibitem{Steiner02}
A.~W. Steiner, S.~Reddy and M.~Prakash,
\newblock Phys. Rev. D {\bf 66}, 094007 (2002).

\bibitem{Schmitt}
K.~Rajagopal and A.~Schmitt,
\newblock Phys. Rev. D {\bf 76}, 045003 (2006).

\bibitem{Shovkovy}
S.~B. Ruester, I.~A. Shovkovy and D.~H. Rischke,
\newblock J. Phys. G {\bf 31}, S849 (2005).

\bibitem{Sharma}
K.~Rajagopal and R.~Sharma,
\newblock Phys. Rev. D {\bf 74}, 094019 (2006).

\bibitem{Neumann}
F.~Neumann, M.~Buballa and M.~Oertel,
\newblock Nucl. Phys. A {\bf 714}, 481 (2003).

\bibitem{Lugones:2002va}
G.~Lugones and J.~E. Horvath,
\newblock Phys. Rev. {\bf D66}, 074017 (2002).

\bibitem{JRS05}
P.~Jaikumar, S.~Reddy and A.~W. Steiner,
\newblock Phys. Rev. Lett {\bf 96}, 041101 (2005).

\bibitem{ARSS06}
M.~G. Alford, K.~Rajagopal, S.~R. Reddy and A.~W. Steiner,
\newblock Phys. Rev. D {\bf 73}, 114016 (2006).

\bibitem{ABPR05}
M.~G. Alford, M.~Braby, M.~W. Paris and S.~R. Reddy,
\newblock Astrophys. J. {\bf 629}, 969 (2005).

\bibitem{Alford:2007xm}
M.~G. Alford, A.~Schmitt, K.~Rajagopal and T.~Schafer,
\newblock (2007).

\bibitem{Shovkovy:2002kv}
I.~A. Shovkovy and P.~J. Ellis,
\newblock Phys. Rev. {\bf C66}, 015802 (2002).

\bibitem{JPS}
P.~Jaikumar, M.~Prakash and T.~Sch\"afer,
\newblock Phys. Rev. D {\bf 66}, 063003 (2002).

\bibitem{Jot}
M.~G. Alford, P.~Jotwani, C.~Kouvaris, J.~Kundu and K.~Rajagopal,
\newblock Phys. Rev. D {\bf 71}, 114011 (2005).

\bibitem{Blaschke1}
H.~Grigorian, D.~Blaschke and D.~Voskresensky,
\newblock Phys. Rev. C {\bf 71}, 045801 (2005).

\bibitem{Blaschke2}
S.~Popov, H.~Grigorian and D.~Blaschke,
\newblock Phys. Rev. C {\bf 74}, 025803 (2006).

\bibitem{JRS06}
P.~Jaikumar, C.~D. Roberts and A.~Sedrakian,
\newblock Phys. Rev. C {\bf 73}, 042801 (2006).

\bibitem{Andy}
N.~Andersson,
\newblock Astrophys. J. {\bf 502}, 708 (1998).

\bibitem{Friedman}
J.~L. Friedman and S.~M. Morsink,
\newblock Astrophys. J. {\bf 502}, 714 (1998).

\bibitem{Morsink}
L.~Lindblom, B.~J. Owen and S.~M. Morsink,
\newblock Phys. Rev. Lett. {\bf 80}, 4843 (1998).

\bibitem{Arras03}
P.~Arras {\em et~al.},
\newblock Astrophys. J. {\bf 591}, 1129 (2003).

\bibitem{AK}
N.~Andersson and K.~D. Kokkotas,
\newblock Int. J. Mod. Phys. {\bf 10}, 381 (2001).

\bibitem{ARZ}
Z.~Arzoumanian, J.~M. Cordes and D.~F. Chernoff,
\newblock Astrophys. J. {\bf 568}, 289 (2002).

\bibitem{Kaspi}
C.-A. Faucher-Giguere and V.~Kaspi,
\newblock Astrophys. J. {\bf 643}, 355 (2006).

\bibitem{Stella}
R.~Perna, R.~Soria, D.~Pooley and L.~Stella,
\newblock arXiv:0712.1040 [astro-ph].

\bibitem{Ott06}
C.~D. Ott, A.~Burrows, T.~A. Thompson, E.~Livne and R.~Walder,
\newblock Astrophys. J. Supp. {\bf 164}, 130 (2006).

\bibitem{Lorimer}
D.~L. Lorimer,
\newblock Living Reviews in Relativity {\bf 8}, 7 (20015).

\bibitem{Lamb}
F.~K. Lamb and S.~Boutloukos,
\newblock arXiv:0705.0155 [astro-ph].

\bibitem{Hessels06}
J.~W.~T. Hessels {\em et~al.},
\newblock Science {\bf 311}, 1901 (2006).

\bibitem{Kaaret07}
P.~Kaaret {\em et~al.},
\newblock Astrophys. J. Lett. {\bf 657}, 97 (2007).

\bibitem{Lars}
L.~Bildsten,
\newblock ApJ {\bf 501}, L89 (1998).

\bibitem{Ben}
B.~J. Owen,
\newblock Phys. Rev. Lett. {\bf 95}, 211101 (2005).

\bibitem{Sterg}
N.~Stergioulas,
\newblock Living Reviews in Relativity {\bf 6}, 3 (2003).

\bibitem{Provost}
J.~Provost, G.~Berthomeiu and A.~Rocca,
\newblock Living Reviews in Relativity {\bf 94}, 126 (1981).

\bibitem{Saio}
H.~Saio,
\newblock Astrophys. J. {\bf 256}, 717 (1982).

\bibitem{PP}
J.~Papaloizou and J.~E. Pringle,
\newblock Mon. Not. R. Astr. Soc. {\bf 182}, 423 (1978).

\bibitem{Chandrasekhar70}
S.~Chandresekhar,
\newblock Phys. Rev. Lett. {\bf 24}, 611 (1970).

\bibitem{Friedman78}
J.~L. Friedman and B.~F. Schutz,
\newblock Astrophys. J. {\bf 222}, 881 (1978).

\bibitem{Mendell}
L.~Lindblom and G.~Mendell,
\newblock Phys. Rev. D {\bf 61}, 104003 (2000).

\bibitem{LMO}
L.~Lindblom, G.~Mendell and B.~J. Owen,
\newblock Phys. Rev. D {\bf 60}, 064006 (1999).

\bibitem{Shapiro:1984}
S.~L. Shapiro and S.~A. Teukolsky,
\newblock {\em Black Holes, White Dwarfs and Neutron Stars} (John Wiley $\&$
  Sons, 1984).

\bibitem{Akmal98}
A.~Akmal, V.~R. Pandharipande and D.~G. Ravenhall,
\newblock Phys. Rev. C {\bf 58}, 1804 (1998).

\bibitem{Wiringa}
R.~B. Wiringa, V.~G.~J. Stoks and R.~Schiavilla,
\newblock Phys. Rev. C {\bf 51}, 38 (1995).

\bibitem{Pudliner}
B.~S. Pudliner, V.~R. Pandharipande, J.~Carlson and R.~B. Wiringa,
\newblock Phys. Rev. Lett {\bf 74}, 4396 (1995).

\bibitem{Forest}
J.~L. Forest, V.~R. Pandharipande and J.~L. Friar,
\newblock Phys. Rev. C {\bf 52}, 568 (1995).

\bibitem{Rajagopal00}
K.~Rajagopal and F.~Wilczek,
\newblock Phys. Rev. Lett. {\bf 86}, 3492 (2000).

\bibitem{ARRW}
M.~G. Alford, K.~Rajagopal, S.~Reddy and F.~Wilczek,
\newblock Phys. Rev. D {\bf 64}, 074017 (2001).

\bibitem{Schafer:1999pb}
T.~Schafer and F.~Wilczek,
\newblock Phys. Rev. {\bf D60}, 074014 (1999).

\bibitem{Fraga01}
E.~S. Fraga, R.~D. Pisarski and J.~Schaffner-Bielich,
\newblock Phys. Rev. D {\bf 63}, 121702 (2001).

\bibitem{Cutler87}
C.~Cutler and L.~Lindblom,
\newblock Astrophys. J. {\bf 314}, 234 (1987).

\bibitem{Flowers79}
E.~Flowers and N.~Itoh,
\newblock Astrophys. J. {\bf 230}, 847 (1979).

\bibitem{Schwenk04}
A.~Schwenk and B.~Friman,
\newblock Phys. Rev. Lett {\bf 92}, 082501 (2004).

\bibitem{TT93}
T.~Takatsuka and R.~Tamagaki,
\newblock Prog. Theor. Phys, Suppl. {\bf 112}, 27 (1993).

\bibitem{Yakovlev01}
D.~G. Yakovlev, A.~D. Kaminker, O.~Y. Gnedin and P.~Haensel,
\newblock Phys. Rept. {\bf 354}, 1 (2001).

\bibitem{Haensel:2001}
P.~Haensel, K.~P. Levenfish and D.~G. Yakovlev,
\newblock Astronomy and Astrophysics {\bf 372}, 130 (2001).

\bibitem{Bedaque:2003wj}
P.~F. Bedaque, G.~Rupak and M.~J. Savage,
\newblock Phys. Rev. {\bf C68}, 065802 (2003).

\bibitem{Alpar}
M.~A. Alpar and J.~A. Sauls,
\newblock Astrophys. J. {\bf 327}, 725 (1988).

\bibitem{madsen}
J.~Madsen,
\newblock Phys. Rev. D {\bf 46}, 3290 (1992).

\bibitem{Basil}
B.~Sa'd, I.~A. Shovkovy and D.~H. Rischke,
\newblock Phys. Rev. D {\bf 75}, 125004 (2007).

\bibitem{Haensel89}
P.~Haensel and A.~J. Jerzak,
\newblock Acta. Phys. Pol. {\bf B20}, 141 (1989).

\bibitem{Son:1999cm}
D.~T. Son and M.~A. Stephanov,
\newblock Phys. Rev. {\bf D61}, 074012 (2000),
\newblock Erratum:~\Journal{\PRD}{62}{059902}{2000}.

\bibitem{Manuel:2007pz}
C.~Manuel and F.~J. Llanes-Estrada,
\newblock JCAP {\bf 0708}, 001 (2007).

\bibitem{Alford:2007pj}
M.~G. Alford and A.~Schmitt,
\newblock AIP Conf. Proc. {\bf 964}, 256 (2007).

\bibitem{Alf}
M.~G. Alford, M.~Braby, S.~Reddy and T.~Sch\"afer,
\newblock Phys. Rev. C {\bf 75}, 055209 (2007).

\bibitem{Khalatnikov}
I.~M. Khalatnikov,
\newblock {\em An Introduction to the Theory of Superfluidity} (Benjamin, New
  York, 1965).

\bibitem{Landau}
E.~M. Lifshitz and L.~P. Pitaevskii,
\newblock {\em Physical Kinetics} (Elsevier, 1987).

\bibitem{Lattimer:1994}
J.~M. Lattimer, K.~A. Van~Riper, M.~Prakash and M.~Prakash,
\newblock Astrophys J. {\bf 425}, 802 (1994).

\bibitem{Schafer:2002ty}
T.~Schafer,
\newblock Phys. Rev. {\bf D65}, 094033 (2002).

\bibitem{Alford:2008pb}
M.~G. Alford, M.~Braby and A.~Schmitt,
\newblock arXiv:0806.0285 [nucl-th].

\bibitem{CM}
C.~Manuel, A.~Dobado and F.~J. Llanes-Estrada,
\newblock Jour. High Energy Phys. {\bf 0509}, 076 (2005).

\bibitem{Zheng06}
X.~Zheng {\em et~al.},
\newblock New Astron. {\bf 12}, 165 (2006).

\bibitem{LOM}
L.~Lindblom, B.~J. Owen and S.~M. Morsink,
\newblock Phys. Rev. Lett. {\bf 80}, 4843 (1998).

\bibitem{Brown02}
E.~F. Brown, L.~Bildsten and P.~Chang,
\newblock Astrophys. J. {\bf 574}, 920 (2002).

\bibitem{Parenti07}
A.~Drago, G.~Pagliara and I.~Parenti,
\newblock arXiv:0704.0510 [astro-ph].

\bibitem{Bildsten00}
L.~Bildsten and G.~Ushomirsky,
\newblock Astrophys. J. Lett. {\bf 529}, 33 (2000).

\bibitem{Levin01}
Y.~Levin and G.~Ushomirsky,
\newblock Mon. Not. R. Astron. Soc. {\bf 324}, 917 (2001).

\bibitem{Glampedakis06}
K.~Glampedakis and N.~Andersson,
\newblock Mon. Not. R. Astron. Soc. {\bf 371}, 1311 (2006).

\bibitem{Lin07}
L.-P. Lin, N.~Andersson and G.~L. Comer,
\newblock arXiv:0709.0660 [gr-qc] .

\end{thebibliography}

\end{document}